\documentclass{aastex}
\usepackage{spr-astr-addons}
\usepackage{url}
\usepackage[utf8]{inputenc}

\RequirePackage{color}

\begin{document}
\title{Charged strange star in $f(R,T)$ gravity with linear equation of state}

\shorttitle{Charged strange star in $f(R,T)$ gravity with linear equation of state}

\shortauthors{Pramit Rej \& Piyali Bhar }

\author{Pramit Rej}

\altaffiltext{}{Department of
Mathematics, Sarat Centenary College, Dhaniakhali, Hooghly, West Bengal 712 302, India }
\altaffiltext{}{Email: pramitrej@gmail.com}

\author{Piyali Bhar }

\altaffiltext{}{ Department of
Mathematics,Government General Degree College,Singur, West Bengal 712409,
India}
\altaffiltext{}{Email:piyalibhar90@gmail.com}

\begin{abstract}
Our present study involves the strange stars model in the framework of $f(R,T)$ theory of gravitation. We have taken a linear function of the Ricci scalar $R$ and the trace $T$ of the stress-energy tensor $T_{\mu \nu}$ for the expression of $f(R,T)$, i.e., $f(R,T)=R+ 2 \gamma T $ to obtain the proposed model, where $\gamma$ is a coupling constant. Moreover, to solve the
hydrostatic equilibrium equations, we consider a linear equation of state between the radial pressure $p_r$ and matter density $\rho$ as $p_r=\alpha \rho-\beta$, where $\alpha$ and $\beta$ are some positive constants,  Both $\alpha,\,\beta$ depend on coupling constant $\gamma$ which have been also depicted in this paper. By employing the Krori-Barua {\em ansatz} already reported in the literature [J. Phys. A, Math. Gen. 8:508, 1975] we have found the solutions of the field equations in  $f (R, T )$ gravity. The effect of coupling constant $\gamma$ have been studied on the model parameters like density, pressures, anisotropic factor, compactness, surface redshift, etc. both numerically and graphically. A suitable range for $\gamma$ is also obtained. The physical acceptability and stability of the stellar system have been tested by different physical tests, e.g., the causality condition, Herrera cracking concept,
relativistic adiabatic index, energy conditions, etc. One can regain the solutions in Einstein gravity when $\gamma\rightarrow 0$.
\end{abstract}



\maketitle

\section{Introduction}


Recently, the LIGO/Virgo Collaboration announced the observation of a merger of a black hole with
mass $23.2_{-1.0}^{+1.1}~M_{\odot}$ with a compact object with mass $2.59^{+0.08}_{-0.09}~M_{\odot}$ \cite{abbott}, where the mass of the secondary component lies within the so-called low mass gap \cite{l1,l2,l3}. Theoretical and observational evidence suggests that black holes of mass less than $5M_{\odot}$ may not be produced by stellar evolution \cite{l2,l3,l4}. According to some candidate equations of state, a stable neutron star must have a mass of at most $3M_{\odot}$ \cite{n1,n2,n3,n4}. If the mass exceeds this limit, it is hypothesized that neutrons lose their individuality
under extreme pressure and breakdown into quarks. A quark star is smaller in size but ultra-dense as compared
to the neutron star. However, increased pressure in its core stops quark stars from collapsing into black
holes. Moreover, estimates of radii of some stellar objects (LMC X-4, 4U 1820-30, Her X-1, etc.)
suggest that their structure and characteristics may be similar to that of strange quark stars.
On the other hand, the relatively small tidal deformability measured in gravitational-wave signal GW170817 do not favor such large values of $M_{\text{max}}$ but rather suggest it is of the order of $2.5M_{\odot}$ \cite{ab1,ab2}. The heaviest neutron star observed to date has a mass of $2.01\pm 0.04~M_{\odot}$ \cite{13}, and the existence of compact objects in the mass regime $[2.5,\,5]M_{\odot}$ is highly uncertain. \par

From the pioneering work done by \cite{rud} it was already proposed that celestial bodies under certain conditions may become anisotropic.
The author observed
that relativistic particle interactions in a very dense nuclear matter medium could lead
to the formation of anisotropies. Indeed, anisotropies arise in many scenarios of a
dense matter medium, like phase transitions \cite{p1}, pion condensation \cite{p2}, or in the
presence of type 3A super-fluid \cite{p3}. The first extensive study on the effect of
pressure anisotropy on the structure of such massive objects
in the framework of general relativity was made by \cite{bl}. The importance and physical reasons
for the inclusion of anisotropic pressure in the solution may be
found in the works of \cite{her} and \cite{iva}. \cite{dg} demonstrated that
pressure anisotropy affects the physical properties, stability
and structure of stellar matter. They proposed that the stability of stellar
bodies is improved for positive measure of anisotropy when
compared to configurations of isotropic stellar objects. In their successive works \cite{dev1,dev2} proved that the presence of anisotropic
pressures enhances the stability of the configuration under
radial adiabatic perturbations as compared to isotropic matter. \cite{Zubair:2019fco} have discussed the anisotropic matter configuration to explore the existence of realistic stellar objects in non-conservative theory named as Rastall theory of gravity. We have assumed a static spherically symmetric metric with linear equation of state (EoS) to formulate the dynamical equations. \cite{Coley:2019tyx} investigated the field equations in the Einstein-aether theory for static spherically symmetric spacetimes and a perfect fluid source and subsequently with the addition of a scalar field with an exponential self-interacting potential. \cite{Ahmad:2018qgb} showed that without the assumption of dark energy it is not possible that perfect fluid spherical gravitational collapse will occur.  The author solved the field equations by assuming linear equation of state  in metric f(R) gravity.\par

At the astrophysical level scales, some observational issues are also persistent. To address these
issues, such as the dark sector of the universe, in current years, several
modified theories of gravity have been introduced. A few of them are $f(R),\, f (T),\,f(G),\, f(R,T),\,f(G,T)$ gravity, etc. The simplest way to extend GR is through $f(R)$ gravity \cite{t14,t15}, for which
$f(R)$ is a general function of the Ricci scalar $R$. Such a theory has successfully described the acceleration of the universe expansion with no need for dark energy \cite{c16,c17,c18}. In the year $2011$, \cite{harko11} initially proposed the concept of $f(R,T)$ gravity by considering an
extension of standard general relativity, where the gravitational Lagrangian
is given by an arbitrary function of the Ricci scalar $R$ and of
the trace $T$ of the stress-energy tensor. However, three different combinations of the functions of $R$ and $T$ were proposed by Harko and his collaborators, the functionals of $f(R, T)$ are: (i)~$f(R, T) = R + 2f(T)$,
(ii)~$f(R, T) = f(R) + f(T)$ and (iii)~$f(R, T) = f(R) + g(R)f(T)$, where $f(R),\,g(R)$
and $f(T)$ are some arbitrary functions of $R$ and $T$. \cite{harko11} also proposed that the $f(R,T)$ gravity model depends on a source
term, representing the variation of the matter stress-energy
tensor with respect to the metric, and its expression
can be obtained as a function of the matter
Lagrangian $\mathcal{L}_m$ and hence for each choice of $\mathcal{L}_m$
a specific set of field equations can be generated.

There exists a large number of works in literature that have been done in the context of $f(R,T)$ gravity in the field of astrophysics as well as in cosmology. \cite{jamil} have reconstructed some cosmological models in this theory of gravity using the functional form $f(R, T) = R^2 + f(T)$. \cite{az1} studied a cylindrically symmetric self-gravitating dynamical object via complexity factor which is obtained through the orthogonal splitting of Reimann tensor in $f(R,T)$ theory of gravity. The model has been discussed for three different forms of $f(R,T)$. Zubair and his collaborators studied the gravitational collapse and stability constraints for spherical and axial symmetry in $f (R, T )$ gravity \cite{zb1,zb2,zb3}. \cite{aa} described the phenomenon of gravitational lensing in the framework of $f (R, T )$ theory. \cite{mm1} obtained the hydrostatic equilibrium configurations of NSs in the $f(R,T)$ gravity has been investigated from a simple barotropic equation of state (EoS) describing matter inside these objects. \cite{mm2} showed that the $f(R, T)$ gravity might
increase the maximum masses of white dwarfs, getting in touch with some observational data. Accelerating cosmological models have been constructed by \cite{sahu} in  $f(R,T)$ modified
gravity theory at the backdrop of an anisotropic Bianchi type-III universe. Sahoo and his collaborators have reconstructed some $f(R, T)$
cosmological models for anisotropic universes \cite{s1,s2,s3,s4,s5}. \cite{hou} studied a few distinguish cosmological
models in the context of $f (R,T )$ gravity to study a matter influenced
era of expanding Universe. \cite{sahoo} obtained accelerating models in the
framework of $f(R,\,T)$ theory of gravity for an anisotropic Bianchi type III
(BIII) universe. \cite{bhard} have investigated the cosmological bouncing solution in LRS Bianchi-I space-time in framework of $f(R,\,T)$ gravity. \cite{mr1} presented a simple and effective new methodology to build up self-gravitating structures within the framework of $f(R,\,T)$ gravity theory by combining two geometrical schemes, the gravitational decoupling by means of minimal geometric deformation and the embedding technique, specifically the class I grasp. \cite{dag} have studied LRS Bianchi type I cosmological models in $f(R,\,T)$ gravity with tilted observers. \cite{gam} studied the slow-roll approximation of cosmic inflation within the context of $f(R,\,T)$ gravity by choosing a minimal coupling between matter and gravity. \cite{ahmed} used the Karmarkar condition to the spherically symmetric non-static radiating star experiencing dissipative gravitational collapse with a heat flux in the framework of $f(R,\,T)$ gravity.

\par
The present paper deals with the study of
a charged compact star in the framework of the $f(R,T)$ gravity in presence of a linear equation of state (EoS).
The paper is organized as follows: we have discussed the interior space-time in Sect.~\ref{int1}. The basic
equations of $f(R,T)$ gravity in presence of charge are given in Sect.~\ref{sec2}. The solution of the field equations has been obtained
by choosing suitable metric {\em ansatz} and a linear equation of state in Sect. \ref{sec4}. A smooth matching of our interior spacetime to the exterior Reissner-Nordstr\"om line element at the boundary outside the event horizon has been discussed and the expression of the constants $A,\,B$ and $C$ in terms of mass and radius are also obtained in Sect.~\ref{sec5}. In the next section, we have discussed about the compactness factor and surface redshift in modified gravity. The physical behaviors of the model in both modified gravity and Einstein gravity have been shown in Sect.~\ref{sec7}. The next section describes the stability conditions via a different useful test of our present model. Some useful findings and concluding remarks have been given in the final section.

\section{Interior space-time}\label{int1}
In curvature coordinates $(t,\,r,\,\theta,\,\phi)$, the interior geometry of our stellar model is described by the following line element,
\begin{equation}\label{line}
ds^{2}=-e^{\nu}dt^{2}+e^{\lambda}dr^{2}+r^{2}d\Omega^{2},
\end{equation}
where $d\Omega^{2}\equiv \sin^{2}\theta d\phi^{2}+d\theta^{^2}$ and the metric co-efficients $\nu$ and $\lambda$ are purely radial functions and they play an important role to obtain the mass and the redshift
functions, respectively, with radial coordinate range
$0~\leq~r~<\infty$. The spacetime will be asymptotically flat
spacetimes if both $\nu(r)$ and $\lambda(r)$ tends to $0$ as $r~\rightarrow~\infty $.
For our present paper, let us assume the metric co-efficients as,
\begin{eqnarray}\label{elambda}
e^{\lambda}= e^{Ar^2}, e^{\nu}&=&e^{Br^2+C},
\end{eqnarray}
where $B,\,A$ are constants of dimension km$^{-2}$ and $C$ is a dimensionless quantity. These metric potentials were given by \cite{kb} and it was successfully used earlier by several authors to model the compact star in the context of general relativity as well as modified gravity.
We note that $e^{\nu}|_{r=0}=e^{C}>0$ and $e^{\lambda}|_{r=0}=1$, moreover,
$\left(e^{\nu}\right)'=2 B e^{C + B r^2} r ,~~~~ \left(e^{\lambda}\right)'=2 A e^{A r^2} r.$
Therefore, the form of the metric potential chosen here ensures
that the metric function is nonsingular, continuous, and well
behaved in the interior of the star. On a physical basis, this
is one of the desirable features for any well-behaved model.
The main advantage of using this {\em ansatz} is that it produces a physically reasonable and singularity free model.\\
For the matter energy-momentum tensor, $T_{\mu\nu}$, we adopt the
anisotropic fluid form, so that in the comoving frame with four velocity
$u^{\mu}=(e^{-\nu/2},\,0,\,0,\,0)$. Let us assume that, inside the stellar interior, the matter is anisotropic in nature and therefore and the corresponding energy-momentum tensor is given by,
\begin{equation}
T_{\nu}^{\mu}=(\rho+p_r)u^{\mu}u_{\nu}-p_t g_{\nu}^{\mu}+(p_r-p_t)v^{\mu}v_{\nu}
\end{equation}
with $ u^{i}u_{j} =-v^{i}v_j = 1 $ and $u^{i}v_j= 0$. Here the vector $v^{i}$ is the space-like vector and $u_i$ is the fluid 4-velocity and it is orthogonal to $v^{i}$,  where $\rho$ is the matter density, $p_r$ and $p_t$ are respectively the radial and transverse pressure in modified gravity.

\section{Basic field Equations in $f(R,T)$ gravity with charge}\label{sec2}
In this section, we shall describe the basic field equations in $f(R,T)$ theory of gravitation. $f(R,T)$ gravity is an extended form of Einstein's general theory of relativity and the Einstein Hilbert action for this gravity in presence of charge is given
by \cite{harko11},
\begin{eqnarray}\label{action}
S&=&\int \left[\frac{1}{16 \pi} f(R,T) +  \mathcal{L}_m+\mathcal{L}_e\right]\sqrt{-g} d^4 x,
\end{eqnarray}
where $f ( R,T )$ represents the general function of Ricci scalar $R$ and trace $T$ of the energy-momentum tensor $T_{\mu \nu}$, $\mathcal{L}_m$ and  $\mathcal{L}_e$ respectively denote the Lagrangian matter density and Lagrangian for the electromagnetic field with $g = det(g_{\mu \nu}$).\\
Corresponding to action (\ref{action}), the field equations of the $f(R,T)$ gravity is given by the following equation,
\begin{eqnarray}\label{frt}
 \frac{\partial f(R,T)}{\partial R} R_{\mu \nu}-\frac{1}{2} g_{\mu \nu} f(R,T) +(g_{\mu \nu }\Box-\nabla_{\mu}\nabla_{\nu})\frac{\partial f(R,T)}{\partial R}\nonumber\\= 8\pi (T_{\mu \nu}+E_{\mu \nu})-\frac{\partial f(R,T)}{\partial T} (T_{\mu \nu}+  \Theta_{\mu \nu}).\nonumber\\
\end{eqnarray}
Where
$\Box \equiv \frac{1}{\sqrt{-g}}\partial_{\mu}(\sqrt{-g}g^{\mu \nu}\partial_{\nu})$ represents the D'Alambert operator, $\nabla_{\nu}$ represents the covariant derivative associated with the Levi-Civita connection of $g_{\mu \nu}$, $\Theta_{\mu \nu}=g^{\alpha \beta}\frac{\delta T_{\alpha \beta}}{\delta g^{\mu \nu}}$.\\
The stress-energy tensor of matter $T_{\mu \nu}$ is calculated as \cite{landau},
\begin{eqnarray}\label{tmu1}
T_{\mu \nu}&=&-\frac{2}{\sqrt{-g}}\frac{\delta \sqrt{-g}\mathcal{L}_m}{\delta \sqrt{g_{\mu \nu}}},
\end{eqnarray}
and its trace is given by $T=g^{\mu \nu}T_{\mu \nu}$. If the Lagrangian matter density $\mathcal{L}_m$ depends only on $g_{\mu \nu}$, not on its derivatives, eqn.(\ref{tmu1}) becomes, $T_{\mu \nu}= g_{\mu \nu}\mathcal{L}_m-2\frac{\partial \mathcal{L}_m}{\partial g_{\mu \nu}}$.\\
Let us assume $F_{\mu \nu}$ denotes the antisymmetric
electromagnetic field strength tensor, defined by
$F_{\mu \nu}=\frac{\partial A_{\nu}}{\partial x^{\mu}}-\frac{\partial A_{\mu}}{\partial x^{\nu}}$ and the electromagnetic energy-momentum tensor $E_{\mu \nu}$ is defined by,
\begin{eqnarray}
E_{\mu \nu}&=&\frac{1}{4 \pi}\left(F_{\mu}^{\alpha}F_{\nu \alpha}-\frac{1}{4}F^{\alpha \beta}F_{\alpha \beta} g_{\mu \nu}\right).
\end{eqnarray}
Now the electromagnetic field strength tensor $E_{\mu \nu}$ satisfied the Maxwell equations,
\begin{eqnarray}
F^{\mu \nu}_{;\nu}=\frac{1}{\sqrt{-g}}\frac{\partial}{\partial x^{\nu}}(\sqrt{-g}F^{\mu \nu})&=&-4\pi j^{\mu},\label{ta}\\
F_{\mu\nu;\lambda}+F_{\nu \lambda;\mu}+F_{\lambda \mu;\nu}&=&0
\end{eqnarray}
where $A_{\nu}=(\phi(r),\,0,\,0,\,0)$ is the four-potential and  $j^{\mu}$ is the
four-current vector, defined by
$j^{\mu}=\frac{\rho_e}{\sqrt{g_{00}}}\frac{dx^{\mu}}{dx^0}$, where $\rho_e$ denotes the proper charge density. For a static matter distribution, the only non-zero component of the four-current is $j^0$. The only two non-vanishing components of
the electromagnetic field tensor $F^{01}$ and $F^{10}$ are related by
$F^{01} = -F^{10}$.
From Eq. (\ref{ta}) the expression for the electric field can be obtained as,
\begin{eqnarray}
F^{01}&=&-e^{\frac{\lambda+\nu}{2}}\frac{q(r)}{r^2},
\end{eqnarray}
where, $q(r)= 4\pi \int_0^r \rho_e e^{\frac{\lambda}{2}} r^2 dr$, and it represents the net charge inside a sphere of radius $`r'$.\\
Now by taking the covariant divergence of (\ref{frt}), the divergence of the stress-energy tensor $T_{\mu \nu}$ can be obtained as, (For details see ref \cite{harko11,hi,farri})
\begin{eqnarray}\label{conservation}
\nabla^{\mu}T_{\mu \nu}&=&\frac{f_T(R,T)}{8\pi-f_T(R,T)}\Big[(T_{\mu \nu}+\Theta_{\mu \nu})\nabla^{\mu}\ln f_T(R,T)\nonumber\\&&+\nabla^{\mu}\Theta_{\mu \nu}-\frac{1}{2}g_{\mu \nu}\nabla^{\mu}T-\frac{8\pi}{f_T}\nabla^{\mu}E_{\mu \nu}\Big].
\end{eqnarray}
One can check, from eqn.(\ref{conservation}), $\nabla^{\mu}T_{\mu \nu}\neq 0$ for $f_T(R,T)\neq 0.$ Therefore, it can be concluded that the system will not be conserved like Einstein gravity.\par
The matter Lagrangian density $\mathcal{L}_m$ could be a function of both density and pressure, i.e., $\mathcal{L}_m = \mathcal{L}_m (\rho, p)$, or it may be an arbitrary function of the density of the matter $\rho$ only, so that $\mathcal{L}_m = \mathcal{L}_m (\rho)$ \cite{har1}. The matter Lagrangian density can be taken as $\mathcal{L}_m=\rho$ in our current paper and the expression of $\Theta_{\mu \nu}=-2T_{\mu \nu}-pg_{\mu\nu}.$\\
To discuss the coupling effects of matter and curvature components in
$f (R, T )$ gravity, we consider
\begin{eqnarray}\label{e}
f(R,T)&=& R+2 \gamma T,
\end{eqnarray}
as proposed by \cite{harko11}, where $\gamma$ is some small positive constant. The term $2\gamma T$ induces time-dependent coupling (interaction) between curvature and
matter. It also corresponds to $\Lambda$ CDM model with a time-dependent cosmological constant \cite{sharif1}.

\section{Our present model in f(R,T) gravity}\label{sec4}

The field equations in $f(R,T)$ gravity is given by,
\begin{eqnarray}
G_{\mu \nu}&=&8\pi (T_{\mu \nu}^{\text{eff}}+E_{\mu \nu}),
\end{eqnarray}
where $G_{\mu \nu}$ is the Einstein tensor and
\begin{eqnarray}
T_{\mu \nu}^{\text{eff}}&=& T_{\mu \nu}+\frac{\gamma}{8\pi}T g_{\mu \nu}+\frac{\gamma}{4\pi}(T_{\mu \nu}-\rho g_{\mu \nu}),
\end{eqnarray}
For the line element (\ref{line}), the field equations in modified gravity can be written as,
\begin{eqnarray}
8\pi\rho^{\text{eff}}+E^2&=&\frac{\lambda'}{r}e^{-\lambda}+\frac{1}{r^{2}}(1-e^{-\lambda}),\label{f1}\\
8 \pi p_r^{\text{eff}}-E^2&=& \frac{1}{r^{2}}(e^{-\lambda}-1)+\frac{\nu'}{r}e^{-\lambda},\label{f2} \\
8 \pi p_t^{\text{eff}}+E^2&=&\frac{1}{4}e^{-\lambda}\left[2\nu''+\nu'^2-\lambda'\nu'+\frac{2}{r}(\nu'-\lambda')\right]. \nonumber\\\label{f3}
\end{eqnarray}

The quantity $q(r)$ actually
determines the electric field as,
\begin{eqnarray}
E(r)&=&\frac{q(r)}{r^2}.
\end{eqnarray}
where $\rho^{\text{eff}}$, $p_r^{\text{eff}}$ and $p_t^{\text{eff}}$ are respectively the density and pressures in Einstein Gravity where
\begin{eqnarray}
\rho^{\text{eff}}&=& \rho+\frac{\gamma}{8\pi}( \rho-p_r-2p_t),\label{r1}\\
p_r^{\text{eff}}&=& p_r+\frac{\gamma}{8\pi}(\rho+3p_r+2p_t),\label{r2}\\
p_t^{\text{eff}}&=& p_t+\frac{\gamma}{8\pi}(\rho+p_r+4p_t).\label{r3}
\end{eqnarray}
The prime denotes differentiation with respect to $`r'$. A
solution of the system of equations (\ref{f1})-(\ref{f3}) will fully specify the gravitational
and thermodynamical behavior of the interior of the stellar object. In the next section, we shall solve these equations under some conditions.\\

To solve the field equations in modified gravity, we consider a linear relationship between the radial pressure and density as,
\begin{eqnarray}\label{s4}
p_r&=& \alpha \rho- \beta,
\end{eqnarray}
where $\alpha$ and $\beta$ are some positive constants whose value depends on the coupling constant $\gamma$ and will be obtained in the coming sections. The Eqn. (\ref{s4}) is a generalized version of the MIT Bag model equation of state.\\
Using the expressions of the metric co-efficients given in (\ref{elambda}), Eqns.~(\ref{f1})-(\ref{f3}), take the following form :
\begin{eqnarray}
8\pi \rho^{\text{eff}}+E^2&=&\frac{1 + e^{-A r^2} (-1 + 2 A r^2)}{r^2},\label{s1}\\
8\pi p_r^{\text{eff}}-E^2&=&\frac{-1 + e^{-A r^2} (1 + 2 B r^2)}{r^2},\label{s2}\\
8 \pi p_t^{\text{eff}}+E^2&=&e^{-A r^2} \left(-A + 2 B + B (-A + B) r^2\right).\nonumber\\\label{s3}
\end{eqnarray}
Adding Eqns. (\ref{s1}) and (\ref{s2}) we get,
\begin{eqnarray}\label{v1}
8\pi \left(\rho^{\text{eff}}+p_r^{\text{eff}}\right)&=& 2(A+B)e^{-Ar^2},
\end{eqnarray}
Solving Eqn. (\ref{v1}), with the help of (\ref{r1}),(\ref{r2}), and (\ref{s4}), the expressions for matter density and pressures in modified gravity are obtained as,
\begin{eqnarray}
\rho &=& \frac{\beta (\gamma + 4 \pi) + (A + B) e^{-A r^2}}{(1 + \alpha)(\gamma + 4 \pi)}, \label{l1}\\
p_r&=& \frac{-\beta (\gamma + 4 \pi) + \alpha(A + B) e^{-A r^2}}{(1 + \alpha)(\gamma + 4 \pi)},\label{l2}\\
p_t &=& \frac{1}{6\gamma + 8 \pi} \bigg[ \frac{2 \beta (\gamma + 4 \pi)}{1 + \alpha}-\frac{1}{r^2}+\frac{e^{-A r^2}}{(1 + \alpha) (\gamma + 4 \pi) r^2}\times\nonumber\\&&\Big[ (1 + \alpha)(\gamma + 4 \pi) + \Big(2 B (\gamma + 4 (2 + \alpha) \pi) \nonumber\\&&- A \big((3 + 5 \alpha) \gamma + 4 (\pi + 3 \alpha \pi)\big)\Big) r^2 \nonumber\\&&- (1 + \alpha) (A - B) B (\gamma + 4 \pi) r^4 \Big] \bigg].\label{l3}
\end{eqnarray}

The expression of electric field $E^2$ in modified gravity is obtained as,
\begin{eqnarray}\label{k4}
E^2&=& \frac{e^{-A r^2}}{(1 + \alpha) (3 \gamma + 4 \pi) r^2} \Big[-2 \gamma -
   2 \alpha \gamma - 4 \pi - 4 \alpha \pi \nonumber\\
&& 	+ 4 A \alpha \gamma r^2 -
   B \gamma r^2 + 3 \alpha B \gamma r^2 + 8 A \alpha \pi r^2 - 8 B \pi r^2 \nonumber\\
&& - A B \gamma r^4 - A \alpha B \gamma r^4 +
   B^2 \gamma r^4 + \alpha B^2 \gamma r^4 \nonumber\\
&& - 2 e^{A r^2} (\gamma + 2 \pi) \big(-1 - \alpha +
      2 \beta (\gamma + 4 \pi) r^2\big)\Big]
\end{eqnarray}
and the expression of anisotropic factor $\Delta=p_t-p_r$ in modified gravity is obtained as,
\begin{eqnarray}
  \Delta &=& \frac{1}{6\gamma + 8 \pi} \bigg[ \frac{2 \beta (\gamma + 4 \pi)}{1 + \alpha}-\frac{1}{r^2}+\frac{e^{-A r^2}}{(1 + \alpha) (\gamma + 4 \pi) r^2}\times\nonumber\\&&\Big[ (1 + \alpha)(\gamma + 4 \pi) + \Big(2 B (\gamma + 4 (2 + \alpha) \pi) \nonumber\\&&- A \big((3 + 5 \alpha) \gamma + 4 (\pi + 3 \alpha \pi)\big)\Big) r^2 \nonumber\\&& - (1 + \alpha) (A - B) B \times (\gamma + 4 \pi) r^4 \Big] \bigg]\nonumber\\&&+\frac{\beta (\gamma + 4 \pi) - \alpha(A + B) e^{-A r^2}}{(1 + \alpha)(\gamma + 4 \pi)}.
\end{eqnarray}
Now we are in a position to obtain the expressions for $\rho^{\text{eff}}$, $p_r^{\text{eff}}$ and $p_t^{\text{eff}}$ which respectively denote the matter density, radial, and transverse pressure in Einstein gravity. The matter density and pressure are thus obtained as,
\begin{eqnarray}
\rho^{\text{eff}}&=& \frac{ 4 \pi\beta + (A + B) e^{-A r^2}}{4 \pi(1 + \alpha)},\label{24}\\
p_r^{\text{eff}}&=& \frac{-4 \pi\beta + \alpha(A + B) e^{-A r^2}}{4 \pi(1 + \alpha)},\label{25}\\
p_t^{\text{eff}}&=& \frac{e^{-A r^2} }{8 (1 + \alpha) \pi r^2}\Big[1 - (A - 4 B) r^2 + B (-A + B) r^4 \nonumber\\&&+
   e^{A r^2} (-1 + 8 \beta \pi r^2) -
   \alpha \big(e^{A r^2} - (1 + B r^2)^2 \nonumber\\&&+ A r^2 (3 + B r^2)\big)\Big].\label{26}
\end{eqnarray}
The expression for the anisotropic factor in general relativity is given by,
\begin{eqnarray}
  \Delta^{\text{eff}} &=&p_t^{\text{eff}}-p_r^{\text{eff}} \nonumber\\
  &=&\frac{e^{-A r^2}}{8 (1 + \alpha) \pi r^2}\Big[1 - A r^2 + 4 B r^2 - A B r^4 \nonumber\\&&+ B^2 r^4 +
   e^{A r^2} (-1 + 16 \beta \pi r^2)\nonumber\\&& -
   \alpha (-1 + e^{A r^2} + 5 A r^2 + A B r^4 - B^2 r^4)\Big].
\end{eqnarray}
So, we have obtained the expressions for the model parameters in the background of General relativity as well as in modified gravity.\\

\begin{figure}[htbp]
    \centering
        \includegraphics[scale=.65]{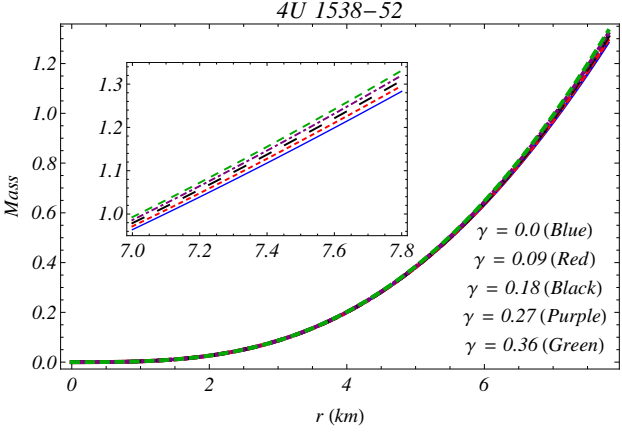}
       \caption{The variation of mass function in $f(R,T)$ gravity is shown against radius for different values of $\gamma$ mentioned in the figure.\label{mass}}
\end{figure}

The gravitational mass within the radius `r' of the charged strange star in modified gravity can be obtained as, \cite{murad}
\begin{eqnarray}\label{m4}
\mathcal{M}&=&4\pi \int_0^r \rho (\tilde{r})~\tilde{r}^2 d\tilde{r} +\frac{q^2}{2r}+\frac{1}{2}\int_0^r\frac{q(\tilde{r})^2}{\tilde{r}^2}d\tilde{r},\nonumber\\
&=& m^{\text{eff}} -\frac{\gamma}{2}\int_0^r(\rho-p_r-2p_t)(\tilde{r})~\tilde{r}^2 d\tilde{r} \nonumber\\&&+\frac{q^2}{2r}+\frac{1}{2}\int_0^r\frac{q(\tilde{r})^2}{\tilde{r}^2}d\tilde{r}.
\end{eqnarray}
Where $m^{\text{eff}}=4\pi \int_0^r \rho^{\text{eff}} (\tilde{r})\tilde{r}^2 d\tilde{r}$ , in equation (\ref{m4}) that represents the mass within the radius $r$ in the uncharged case in the general theory of relativity(GTR). For $\gamma=0$, we obtain the mass of the charged strange star in Einstein's general theory of relativity. From equation (\ref{m4}) we get,
\begin{eqnarray}
\mathcal{M}&=& \frac{r}{2} (1 - e^{-A r^2}) + \frac{re^{-A r^2}}{2 (1 + \alpha) (3 \gamma + 4 \pi)}
    \Big(-2 \gamma - 2 \alpha \gamma \nonumber\\
    &&- 4 \pi - 4 \alpha \pi +
    4 A \alpha \gamma r^2 - B \gamma r^2 + 3 \alpha B \gamma r^2  \nonumber\\
    &&+
    8 A \alpha \pi r^2 - 8 B \pi r^2 - A B \gamma r^4 -
    A \alpha B \gamma r^4 \nonumber\\ && + B^2 \gamma r^4 + \alpha B^2 \gamma r^4
     +2 e^{A r^2} (\gamma + 2 \pi)\times\nonumber\\&&\big(1+\alpha -
       2 \beta (\gamma + 4 \pi) r^2\big)\Big).
\end{eqnarray}
The mass function is regular at the center as $\mathcal{M}~\rightarrow 0$ for $r\rightarrow 0$. The profile of the mass function is plotted in Fig.~\ref{mass}.\par

\section{Matching Condition}\label{sec5}

\cite{jun} obtained the constants $A,\,B,\,C$ of the Krori-Barua metric in terms of the
physical constants of mass, charge, and radius of the source. The author also investigated the conditions for physical
relevance that leads to a functional dependence of the ratio of mass-to-radius
on the ratio of charge-to-mass and also to upper and lower limits on these ratios. Now to be a physically reasonable model, our interior solution should match smoothly to the exterior spacetime outside the event horizon $r>M+\sqrt{M^{2}-Q^2}$, where, $Q$ is the total charge enclosed within the boundary $r=R$ and this matching condition yields the constants $A,\,B$ and $C$. The exterior space-time of the star will be described by the Reissner-Nordstr\"om spacetime \cite{rn1,rn21} given by
\begin{eqnarray}
ds^{2} &=& -\left(1 - \frac{2M}{r} + \frac {Q^2}{r^2}\right)dt^2 + \left(1 - \frac{2M}{r} + \frac {Q^2}{r^2}\right)^{-1}dr^2
\nonumber\\
&& + r^2(d\theta^2+\sin^2\theta d\phi^2), \label{eq22}
\end{eqnarray}
Now a proper junction condition for an isotropic fluid sphere at a stellar surface is that of O'Brien-Synge \cite{ref1,ref2} as follow:
\begin{eqnarray}
\text{\bf{junction condition~1}}: g_{ab}(R+0) = g_{ab}(R-0),\nonumber\\\label{j2}
  \end{eqnarray}
where $g_{ab}$ is the metric components $(a,b=t,\theta, \phi).$ The Eqn. (\ref{j2}) gives,
 \begin{eqnarray}
1 - \frac{2M}{R} + \frac {Q^2}{R^2} &=& e^{BR^2+C},\label{eq23}\\
1 - \frac{2M}{R} + \frac {Q^2}{R^2} &=& e^{-AR^2},\label{eq24}
\end{eqnarray}
Solving the Eq.~(\ref{eq24}), one can determine the values of the constants $A$ as,
\begin{eqnarray}
A &=& - \frac{1}{R^2} \ln \left[ 1 - \frac{2M}{R} + \frac {Q^2}{R^2}
\right], \label{eq26}
\end{eqnarray}
 \begin{eqnarray}
\text{\bf{junction condition~2}}:\partial_r g_{ab}(R+0) = \partial_r g_{ab}(R-0),\nonumber\\\label{j3}
 \end{eqnarray}
$\frac{\partial g_{tt}}{\partial r}$ across the boundary surface $r= R$ between the interior and the exterior regions give the following relation:
\begin{eqnarray}
\frac{M}{R^2} - \frac {Q^2}{R^3} &=& B Re^{BR^2+C}.\label{eq25}
\end{eqnarray}
Solving (\ref{eq23}) and (\ref{eq25}), $B$ and $C$ in terms of the total mass $M$, radius $R$, and charge $Q$ can be obtained as,
\begin{eqnarray}
B &=& \frac{1}{R^2} \left[\frac{M}{R} - \frac {Q^2}{R^2}\right] \left[1 - \frac{2M}{R} + \frac {Q^2}{R^2}
\right]^{-1},\label{eq27}\\
C &=&  \ln \left[ 1 - \frac{2M}{R} +
\frac {Q^2}{R^2} \right]- \frac{ \frac{M}{R} - \frac {Q^2}{R^2}}{
\left[ 1 - \frac{2M}{R} + \frac {Q^2}{R^2} \right]}. \label{eq28}
\end{eqnarray}
\begin{eqnarray}
\text{\bf{junction condition~3}}: p(R+0)= p(R-0),\label{j1}
\end{eqnarray}
The Eqn.(\ref{j1}) changes since we have to take care of both radial and transverse pressures. Now, from Eqn. (\ref{j1}), in presence of anisotropy $p_r(r=R)=0$ gives,
\begin{equation}\label{bn2}
\alpha \rho_s =\beta,
\end{equation}
where $\rho_s$ is the surface density given by,
\begin{eqnarray}
\rho_s&=& \frac{e^{-A R^2} (A + B + \beta e^{A R^2} (\gamma + 4 \pi))}{(1 +
   \alpha) (\gamma + 4 \pi)},
\end{eqnarray}
One can easily check from the profile of $p_t$ given in Fig.~\ref{pr} that the transverse pressure $p_t$ does not vanish at the surface of the star, therefore, there is a discontinuity in tangential pressure. Moreover we have matched our interior spacetime to the exterior Reissner-Nordstr\"om spacetime at the boundary. Obviously the metric coefficients are continuous at the boundary, but it does not ensure that their derivatives are also continuous at the junction surface. In other words the affine connections may be discontinuous there. To take care of this let us use the
Darmois-Israel formation to determine the surface stresses at the junction boundary.
Therefore to avoid the discontinuity the surface stresses at
the junction boundary can be calculated by using the Darmois-\cite{d1,d2} formation.
The surface stress energy ($\sigma$) and surface pressure ($\mathcal{P}$) for our present model are obtained as,
\begin{eqnarray*}
  \sigma &=& -\frac{1}{4\pi R}\Big[\sqrt{1-\frac{2M}{R}+\frac{Q^2}{R^2}}-e^{-\frac{aR^2}{2}}\Big], \\
  \mathcal{P}&=& \frac{1}{8\pi R}\left[\frac{1-\frac{M}{R}}{\sqrt{1-\frac{2M}{R}+\frac{Q^2}{R^2}}}-(1+BR^2)e^{-\frac{aR^2}{2}}\right].
\end{eqnarray*}
{\bf Determination of the constants $\alpha$ and $\beta$~:}\\
We also impose the condition $E^{2}(r=0)=0$ which implies,
\begin{eqnarray}\label{bn1}
2 A (1 + 3 \alpha) (\gamma + 2 \pi) -  B (\gamma - 3 \alpha \gamma + 8 \pi) \nonumber\\=
 4 \beta (\gamma^2 + 6 \gamma \pi + 8 \pi^2),
\end{eqnarray}
Solving (\ref{bn1}) and (\ref{bn2}), one can obtain the expressions of $\alpha$ and $\beta$ in terms of $A,\,B$ and $\gamma$ as,
\begin{eqnarray}
\alpha&=& \frac{e^{A R^2} (-2 A (\gamma + 2 \pi) + B (\gamma + 8 \pi))}{-4 (A + B) (\gamma + 2 \pi) +
 3 e^{A R^2} (B \gamma + 2 A (\gamma + 2 \pi))},
 \label{b7}\nonumber\\
 \\
  \beta&=& \frac{\alpha (A + B) e^{-A R^2}}{\gamma + 4 \pi},\label{b8}
\end{eqnarray}
One can notice that both $\alpha$ and $\beta$ depend on the coupling constant $\gamma$. The variation of $\alpha$ and $\beta$ with respect to $\gamma$ have been discussed both graphically and numerically in the later sections.


\section{Compactness and surface redshift}\label{sec6}
The compactification factor for our present model is obtained as,
\begin{eqnarray}
\mathcal{U}=\frac{\mathcal{M}(R)}{R},
\end{eqnarray}
The compactness factor plays an important role to classify the compact object as, (i) for normal star $\mathcal{U}~\sim~10^{-5}$, (ii) for white dwarfs: $\mathcal{U}~\sim~10^{-3}$ (iii) for neutron star: $10^{-1}~<\mathcal{U}~<\frac{1}{4}$ (iv) for ultra compact star: $\frac{1}{4}~<~\mathcal{U}~<\frac{1}{2}$
(v) for Black hole: $\mathcal{U}=\frac{1}{2}$.\\
Now the bounds for the ratio of mass to the radius of a charged compact star is given by,
\begin{eqnarray}
\frac{3Q^2}{4R^2}\frac{1+\frac{Q^2}{18R^2}}{1+\frac{Q^2}{12R^2}} \leq \mathcal{U} \leq \left(\frac{1}{3}+\sqrt{\frac{1}{9}+\frac{Q^2}{3R^2}}\right)^2.\label{w4}
\end{eqnarray}
In eqn.~(\ref{w4}), $R$ represents the radius of the fluid distribution, $q(r=R)=Q$, where, $Q$ is the total charge inside the fluid sphere. The lower and upper bound are respectively obtained by \cite{harko15} and \cite{an1}. From the eqn. (\ref{w4}), it can be checked that $\mathcal{U}$ obeys the Buchdahl's limit \cite{buch} $2\mathcal{U}<\frac{8}{9}$ for uncharged case.

\begin{table}[t]
\centering
\caption{\label{tab1}The bounds for the inequality (\ref{w4}) for different values of $\gamma$ }
\begin{tabular}{ c | c |c  |c }
\hline
$\gamma$ &  Value of lower & $\mathcal{U}$&  Value of upper\\
&limit of eq.~(\ref{w4})& &limit of eq.~(\ref{w4})\\
\hline
$0.0$ &$0.00749792$&$0.164519$&$0.666717$\\
$0.09$ &$ 0.0435352$&$0.166159$&$0.668353$\\
$0.18$ &$0.0604501$&$0.167732$&$0.669914$ \\
$0.27$ &$0.0730714$&$0.169241$&$0.671405$ \\
$0.36$&$0.0834022$&$0.170690$&$0.672831$\\
\hline
\end{tabular}
\end{table}

Table \ref{tab1} verifies the inequality given in (\ref{w4}) for our present model for different values of the coupling constant $\gamma$.\\

Now the surface redshift for a compact star model is obtained as, $z_s(R)=\frac{1}{\sqrt{\left(1-2\mathcal{U}\right)}}-1$. The numerical values of the surface redshift for different values of $\gamma$ have been presented in Table \ref{tab2}.
\begin{table}[t]
\centering
\caption{\label{tab2}The values of the effective mass, effective compactness and effective surface redshift have been presented for the compact star 4U 1538-52 by assuming $M = 0.87~M_{\odot},\, R = 7.8 $ km., Q = 0.078.}
{\begin{tabular}{@{}cccc@{}}
\hline
$\gamma$&$M^{\text{eff}}$ & $\mathcal{U}$ & $z_s^{\text{eff}}(R)$ \\ \hline
0.0&  1.28325 & 0.1645 & 0.220819\\
0.09 &1.29604 & 0.1662& 0.223814 \\
0.18 & 1.30831& 0.1677 &0.226707\\
0.27& 1.32008& 0.1692 & 0.229502\\
0.36 & 1.33139 & 0.1707& 0.232205\\
\hline
\end{tabular}}
\end{table}

\section{Physical acceptability conditions}\label{sec7}
In this section, we are interested to study more details about the various properties of the stellar configuration by performing
some analytical calculations as well as by drawing the profiles of various model parameters
of obtained solutions and finally, we compare our results with the General relativity (GR) model and the observational
constraints as well.

\begin{figure}[htbp]
    \centering
        \includegraphics[scale=.65]{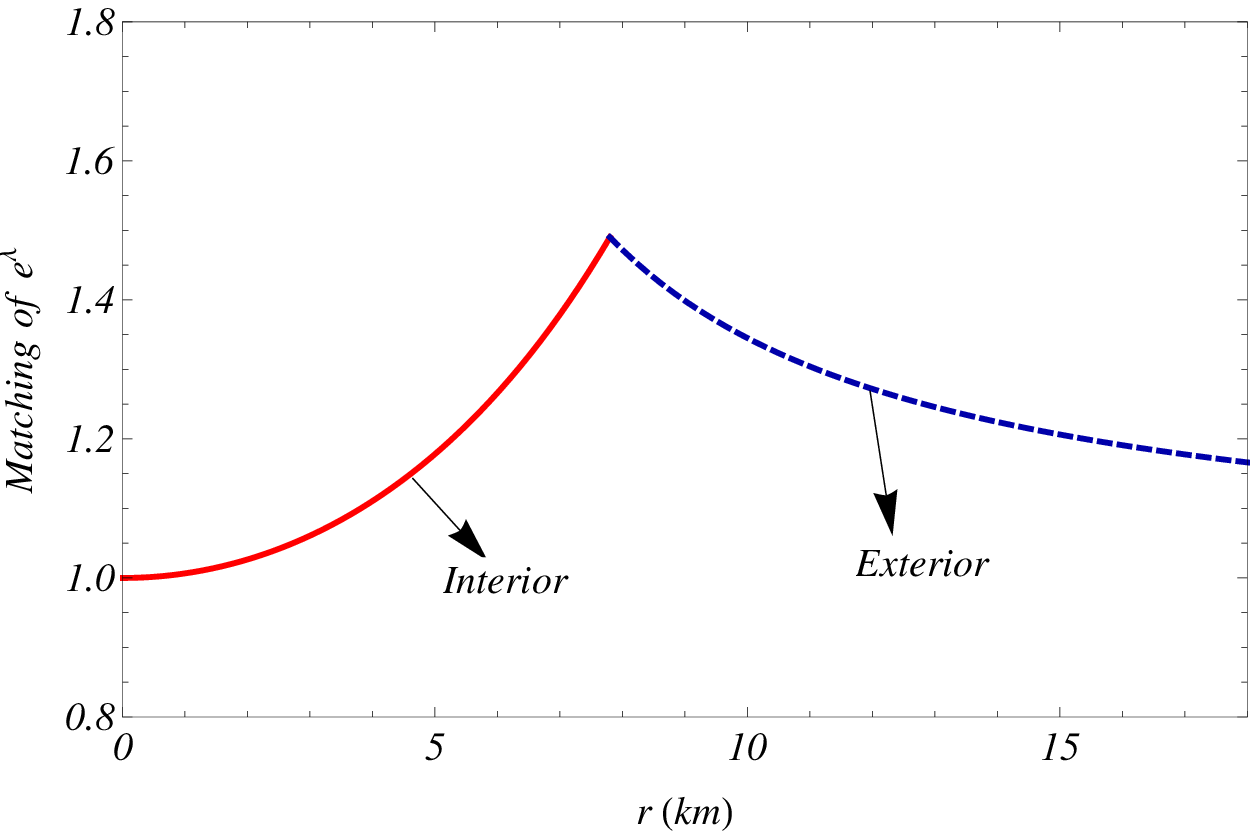}
        \includegraphics[scale=.65]{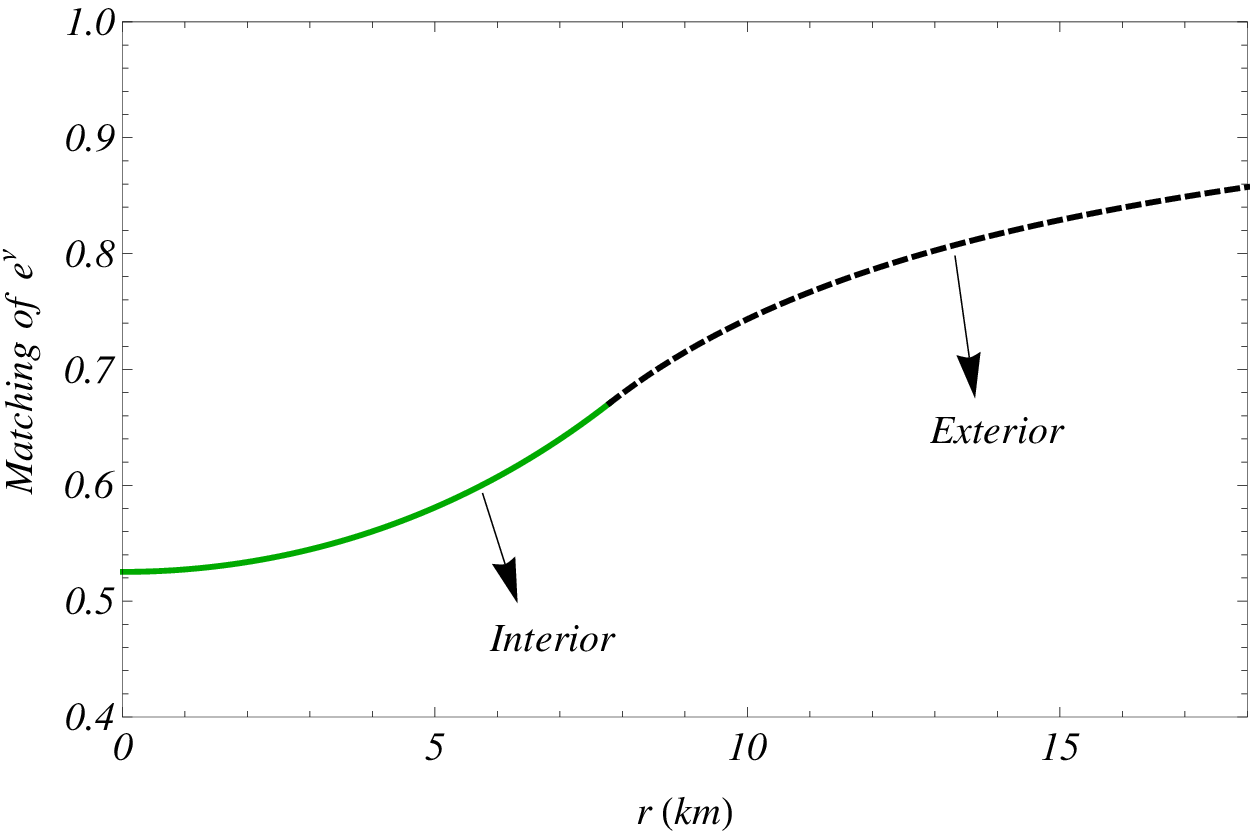}
       \caption{The matching condition of the metric potential $e^{\lambda}$ and $e^{\nu}$ are
shown against radius for the compact star 4U1538-52 by taking the values of
the constants $A,\,B$ and $Q$ mentioned in Table \ref{tab3}.\label{metric}}
\end{figure}

\begin{table*}[t]
\centering
\caption{\label{tab3}The values of the constants $\alpha,\,\beta$, central density, surface density, central pressure and central value of radial adiabatic index have been presented for the compact star 4U 1538-52 by assuming $M = 0.87~M_{\odot},\, R = 7.8 $ km., Q = 0.078.}
{\begin{tabular}{@{}ccccccc@{}}\hline
$\gamma$&$\alpha$&$\beta$&$\rho_c$&$\rho_s$&$p_c$&$\Gamma_{r0}$ \\ \hline
0.0& 0.274101& $1.54916\times 10^{-4}$& $1.05601 \times 10^{15}$& $7.62612\times 10^{14}$ & $7.23777\times 10^{34}$ &3.87337\\
0.09& 0.254518& $1.42826\times 10^{-4}$& $ 1.05305\times 10^{15}$ &$7.57189\times 10^{14}$ & $6.77705\times 10^{34}$ &3.81384 \\
0.18&0.236049& $1.31526\times 10^{-4}$& $ 1.05000\times 10^{15}$&  $ 7.51843\times 10^{14}$ & $6.33414\times 10^{34}$ &3.75769\\
0.27&0.218601& $1.2095\times 10^{-4}$&  $1.04688\times 10^{15}$&  $7.46572\times 10^{14}$ & $5.90823\times 10^{34}$ &3.70465 \\
0.36&0.202093& $1.11038\times 10^{-4}$& $1.04368\times 10^{15}$& $ 7.41374 \times 10^{14}$& $5.49850\times 10^{34}$ &3.65446\\
\hline
\end{tabular} \label{ta1}}
\end{table*}

\subsection{Regularity and Maximality Criterion for pressure and density}
\begin{figure*}[htbp]
    \centering
        \includegraphics[scale=.6]{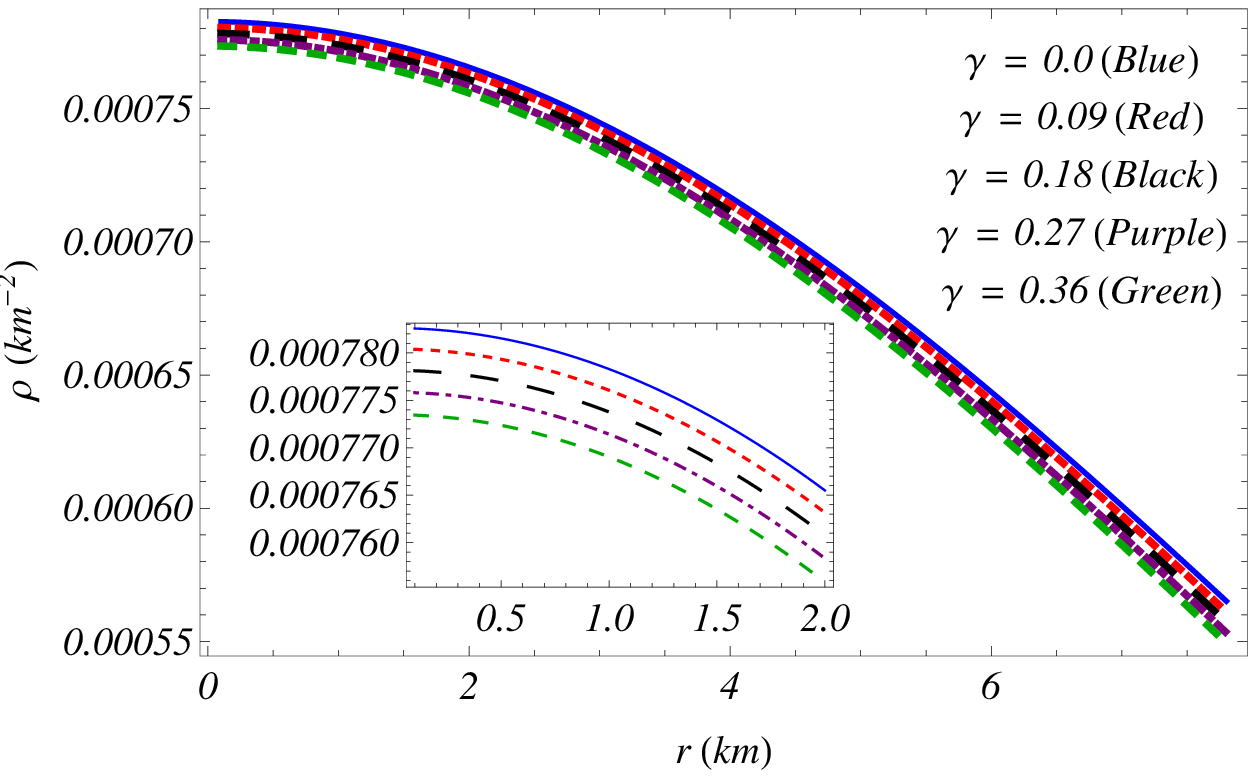}
        \includegraphics[scale=.6]{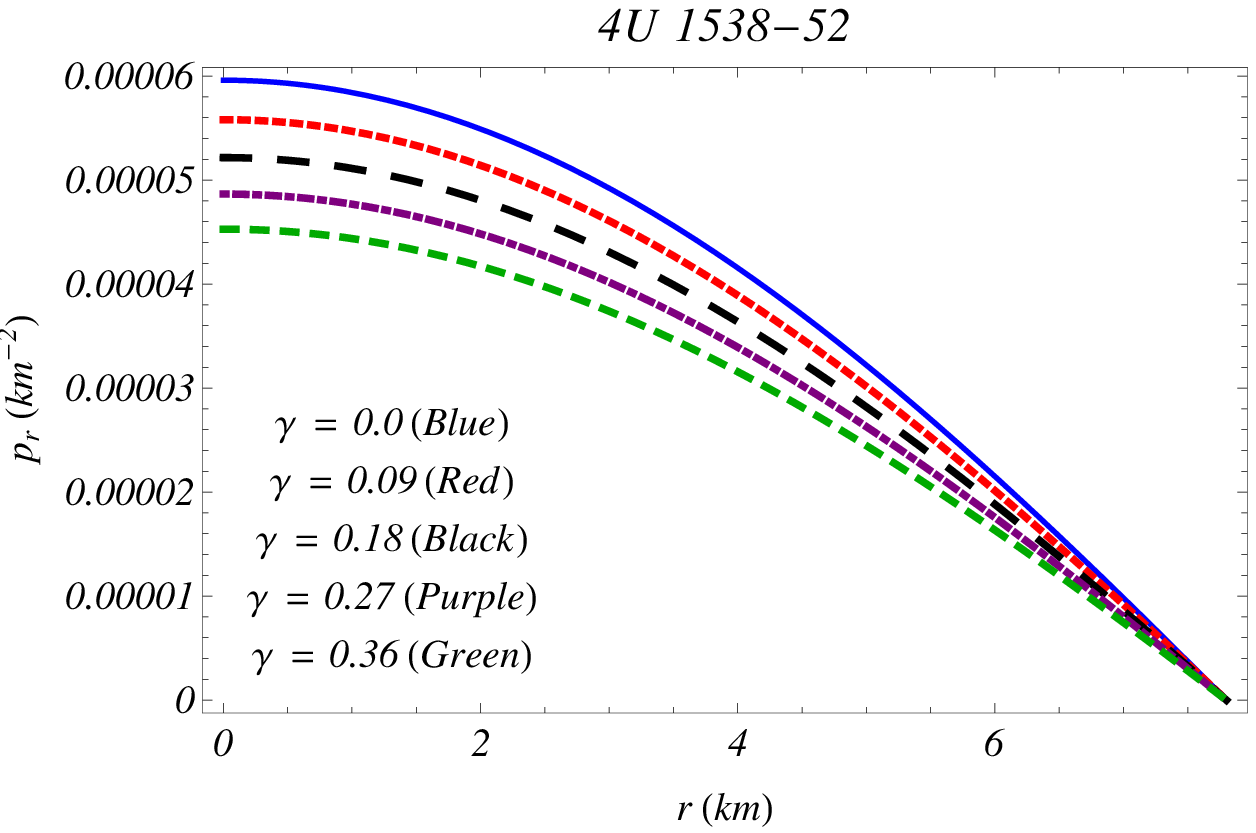}
        \includegraphics[scale=.6]{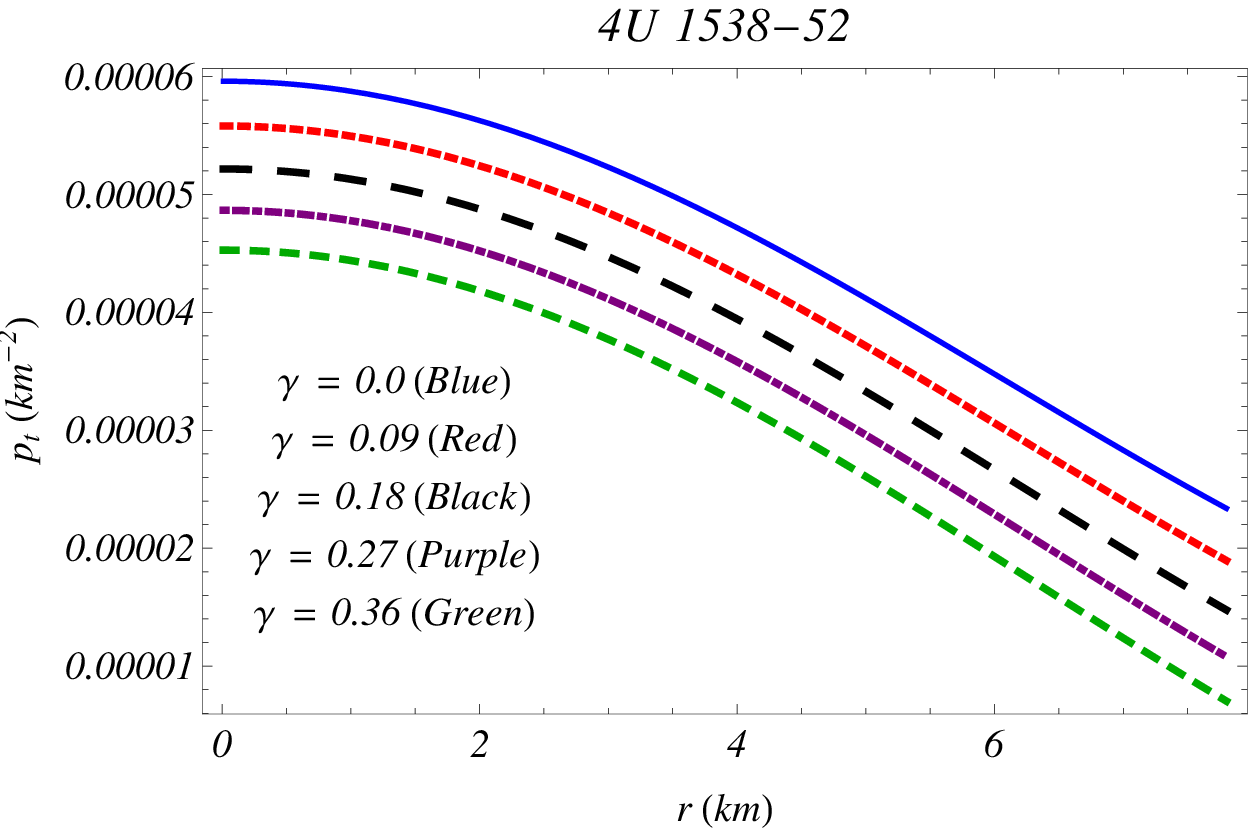}
       \caption{(Top left) Matter density $\rho$, (top right) Radial pressure $p_r$,  (bottom) transverse pressure $p_t$ are plotted against $r$ inside the stellar interior for different values of $\gamma$ mentioned in the figure. \label{pr}}
\end{figure*}

The matter density $\rho$ and both radial and transverse
pressures $p_r,\,p_t $ versus the radial coordinate `r' for the compact star $4U 1538-52$
have been shown in Fig.~\ref{pr} for the numerical values of the parameters mentioned in Table \ref{tab3}.

We can note that the radial pressure vanishes at the surface, neither the transverse pressure nor the matter density vanishes there.
We obtain the density and pressure gradients for our present model by taking the differentiation of the eqns.~(\ref{l1})-(\ref{l3}) with respect to r as,
\begin{eqnarray}
\frac{d\rho}{dr}&=& -\frac{2 A (A + B) e^{-A r^2} r}{(1 + \alpha) (\gamma + 4 \pi)}<0,\label{k1}\\
      \frac{dp_r}{dr}&=& -\frac{2 A \alpha (A + B) e^{-A r^2} r}{(1 + \alpha) (\gamma + 4 \pi)}<0,\label{k2}\\
      \frac{dp_t}{dr}&=&\frac{e^{-A r^2}}{(1 + \alpha) (\gamma + 4 \pi) (3 \gamma + 4 \pi) r^3} \Big[4 \pi \big( h_1(r) \big) \nonumber\\ && + \alpha h_2(r) + \gamma \big(h_2(r) + \alpha h_3(r)\big)\Big]<0.\label{k3}
\end{eqnarray}
where,
\begin{eqnarray*}
h_1(r)&=& (A^2 - 5 A B + B^2) r^4 + h_4(r),\\
h_2(r)&=& (3 A^2 - 3 A B + B^2) r^4 + h_4(r),\\
h_3(r)&=&  (5 A^2 - A B + B^2) r^4 + h_4(r), \\
h_4(r)&=& A (A - B) B r^6-1 + e^{A r^2} - A r^2.
\end{eqnarray*}
We see that \[\left(\frac{d\rho}{dr}\right)|_{r=0}=0,\,\left(\frac{dp_r}{dr}\right)|_{r=0}=0,\,\left(\frac{dp_t}{dr}\right)|_{r=0}=0.\]
and,
\begin{eqnarray*}
\left(\frac{d^2\rho}{dr^2}\right)|_{r=0}&=&-\frac{2 A (A + B)}{(1 + \alpha) (\gamma + 4 \pi)},\\
\left(\frac{d^2p_r}{dr^2}\right)|_{r=0}&=&-\frac{2 A \alpha (A + B)}{(1 + \alpha) (\gamma + 4 \pi)},\\
\left(\frac{d^2p_t}{dr^2}\right)|_{r=0}&=&-\frac{1}{(1 + \alpha) (\gamma + 4 \pi) (6 \gamma + 8 \pi)}\Big[2 B^2  \times\\&& (1 + \alpha)(\gamma + 4 \pi) +
 A^2 \big((7 + 3 \alpha) \gamma \nonumber\\&&+ 12 (1 + \alpha) \pi\big) -
 2 A B \big((3 + 5 \alpha) \gamma \\&&+ 20 (1 + \alpha) \pi \big)\Big].
\end{eqnarray*}
Since the density is a monotonic decreasing function of radius `r', i.e., it has a maximum value at the center of the star, it requires $\left(\frac{d^2\rho}{dr^2}\right)|_{r=0}<0$ and this implies,
\begin{eqnarray}
\gamma+4\pi~>0, \label{w9}
\end{eqnarray}
 since $A,\,B$ and $\alpha$ are all positive.\par
The central density and central pressure for modified gravity are obtained as,
\begin{eqnarray*}
\rho_c&=& \frac{A + B + \beta (\gamma + 4 \pi)}{(1 + \alpha) (\gamma + 4 \pi)},\\
 p_c&=& \frac{\alpha (A + B) -  \beta (\gamma + 4 \pi)}{(1 + \alpha) (\gamma + 4 \pi)}.
\end{eqnarray*}

$p_r(r=0)=p_c>0$ gives, $\alpha (A + B) -  \beta (\gamma + 4 \pi)>0,$ by using (\ref{w9}) and $\alpha>0$.
\begin{eqnarray}\Rightarrow~\frac{\alpha(A+B)}{\beta}-4\pi>\gamma,\label{w1}\end{eqnarray}
Now by \cite{zel} conditions $\frac{p_c}{\rho_c}<1$ gives,
\begin{eqnarray}\frac{(\alpha-1)(A+B)}{2\beta}-4\pi<\gamma,\label{w2}\end{eqnarray}
Combining (\ref{w1}) and (\ref{w2}) we get a range of the coupling constant $\gamma$ as,
\begin{eqnarray}\frac{(\alpha-1)(A+B)}{2\beta}-4\pi<\gamma<\frac{\alpha(A+B)}{\beta}-4\pi.\end{eqnarray}
\subsection{Behavior of electric field and anisotropic factor}
\begin{figure}[htbp]
    \centering
        \includegraphics[scale=.6]{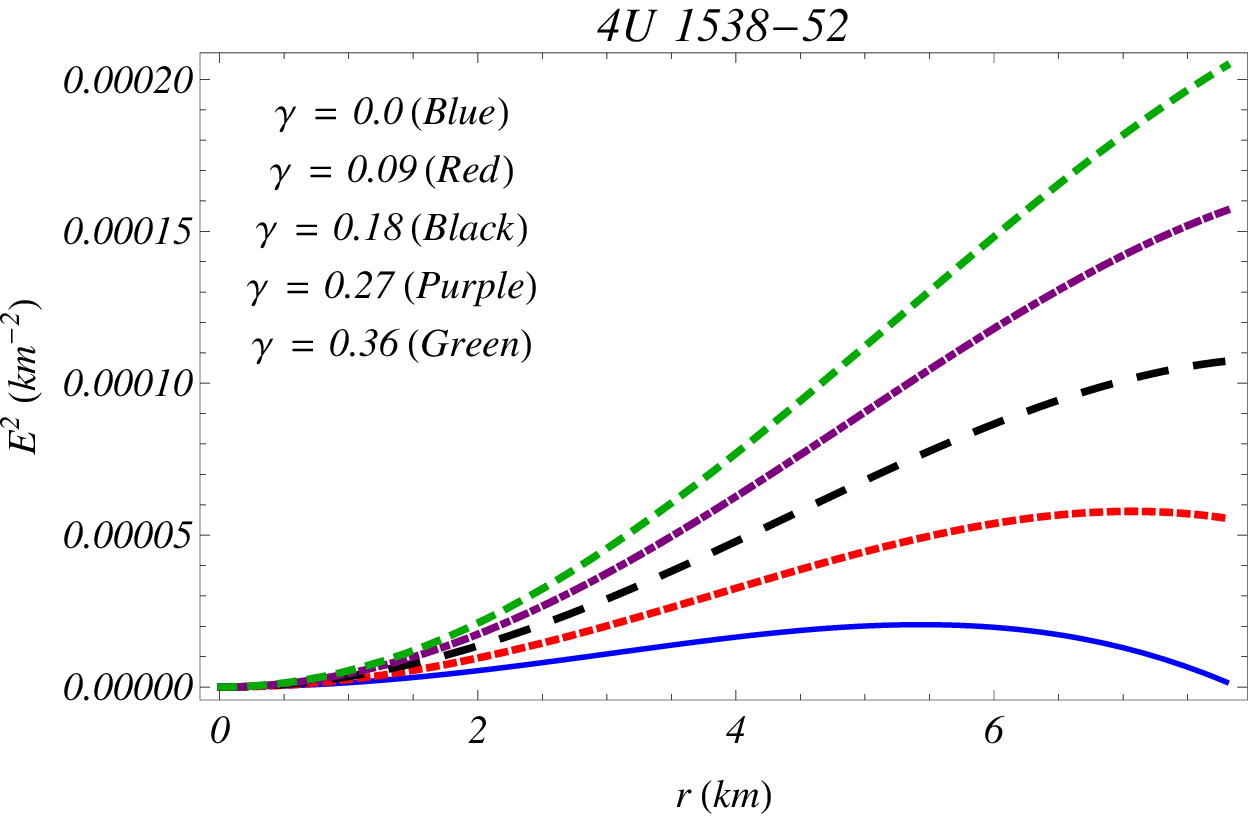}
        \includegraphics[scale=.6]{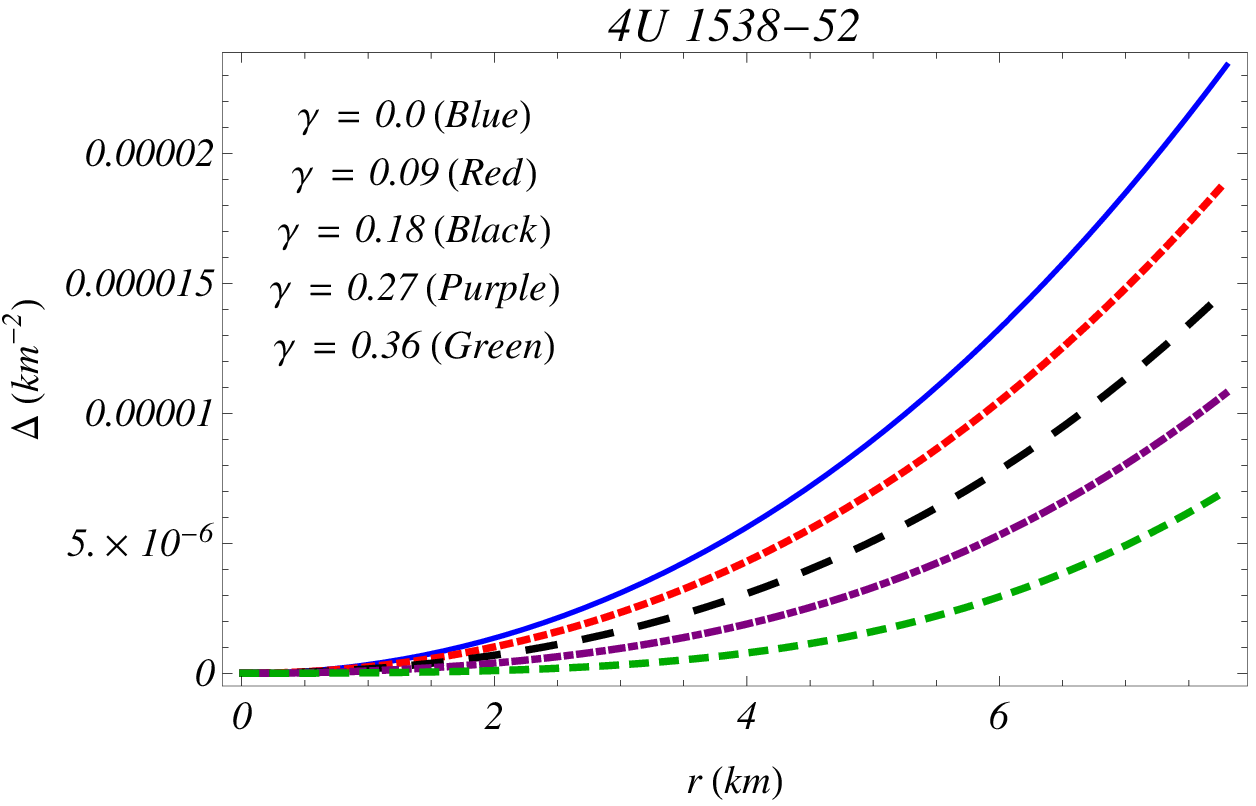}
       \caption{(Left) $E^2$ and (right) anisotropic factor $\Delta$ are shown against radius for different values of $\gamma$ mentioned in the figure.\label{ener}}
\end{figure}

For massive stellar objects, the radial pressure is not
 equal to the tangential one in general. From the pioneering work of \cite{rud}, it is well known that if the density of the core of the star becomes beyond the nuclear density ($\sim~10^{15}~gm./cm^3$), star becomes anisotropy. There are a number of arguments in literature for the existence of the anisotropy in stellar
models such as the existence of a solid core or by
the presence of a type-P superfluid, different kinds of phase transitions, pion condensation, etc. which have been mentioned earlier. In fact,
anisotropy is also important to understand the peculiar
properties of matter in the core of the stellar structure.
The anisotropic factor $\Delta=p_t-p_r$ is shown in Fig.~\ref{ener}. In the current study, the anisotropy function is
positive with a regularly increasing behavior and it vanishes at the center of the star. On the other hand, the behavior of electric field $E^2$ is shown in Fig.~\ref{ener} for different values of $\gamma$. It is evident that the electric field starts from zero at the center and increases towards the surface of the star except for the case $\gamma=0$ and it can be noted that at the surface of the star $E^2|_{r=R}>0,\,\Delta|_{r=R}>0$.

\subsection{ Energy conditions }
All the energy conditions, namely null energy condition (NEC), weak energy condition (WEC), strong energy condition (SEC), and dominant energy condition (DEC) are said to be satisfied for a physically acceptable model if the model parameters $\rho,\,p_r,\,p_t$ and $E^2$ satisfy the following inequalities. It also acts as an important role in understanding the nature of matter \cite{33}.
\begin{itemize}
\item NEC:~$\rho+p_r \geq 0,~\rho + p_t +\frac{E^2}{4\pi} \geq 0,$
\item WEC:~$\rho+p_r \geq 0,~\rho + p_t +\frac{E^2}{4\pi} \geq 0,~ \rho + \frac{E^2}{8\pi} \geq 0,$
\item SEC:~$\rho+p_r \geq 0,~\rho + p_t +\frac{E^2}{4\pi} \geq 0, \rho+ p_r +2 p_t+ \frac{E^2}{4\pi} \geq 0,$
\item DEC:~$\rho-p_r +\frac{E^2}{4 \pi}\geq 0,~\rho - p_t \geq 0,~ \rho + \frac{E^2}{8\pi} \geq 0.$
\end{itemize}

\begin{figure*}[htbp]
    \centering
		\includegraphics[scale=.65]{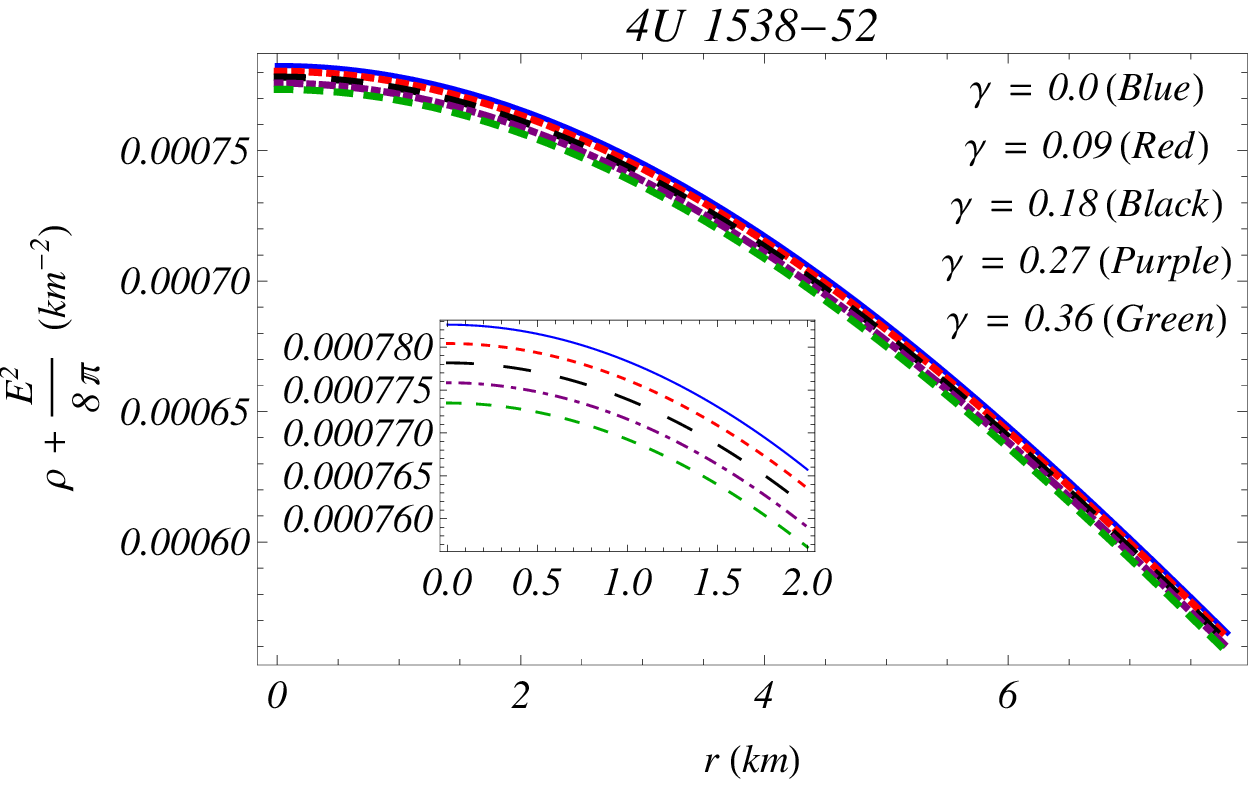}
        \includegraphics[scale=.65]{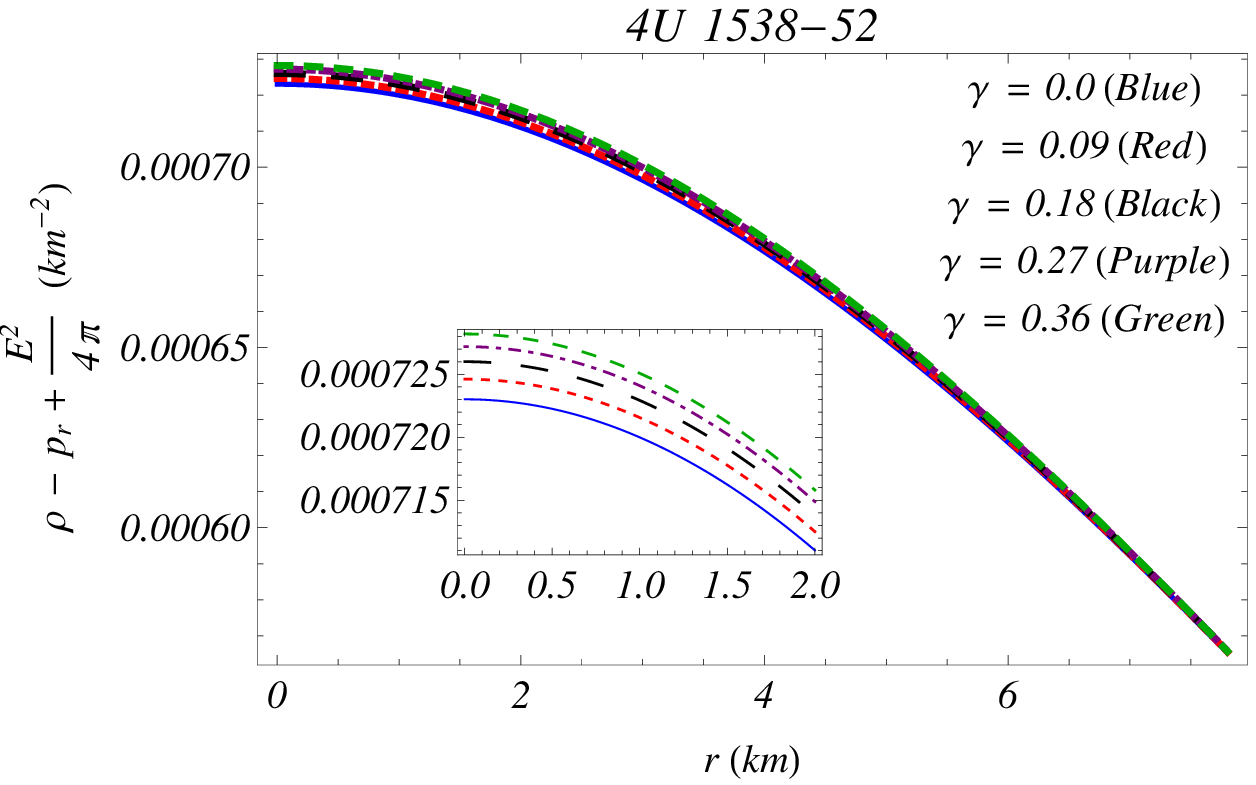}
        \includegraphics[scale=.65]{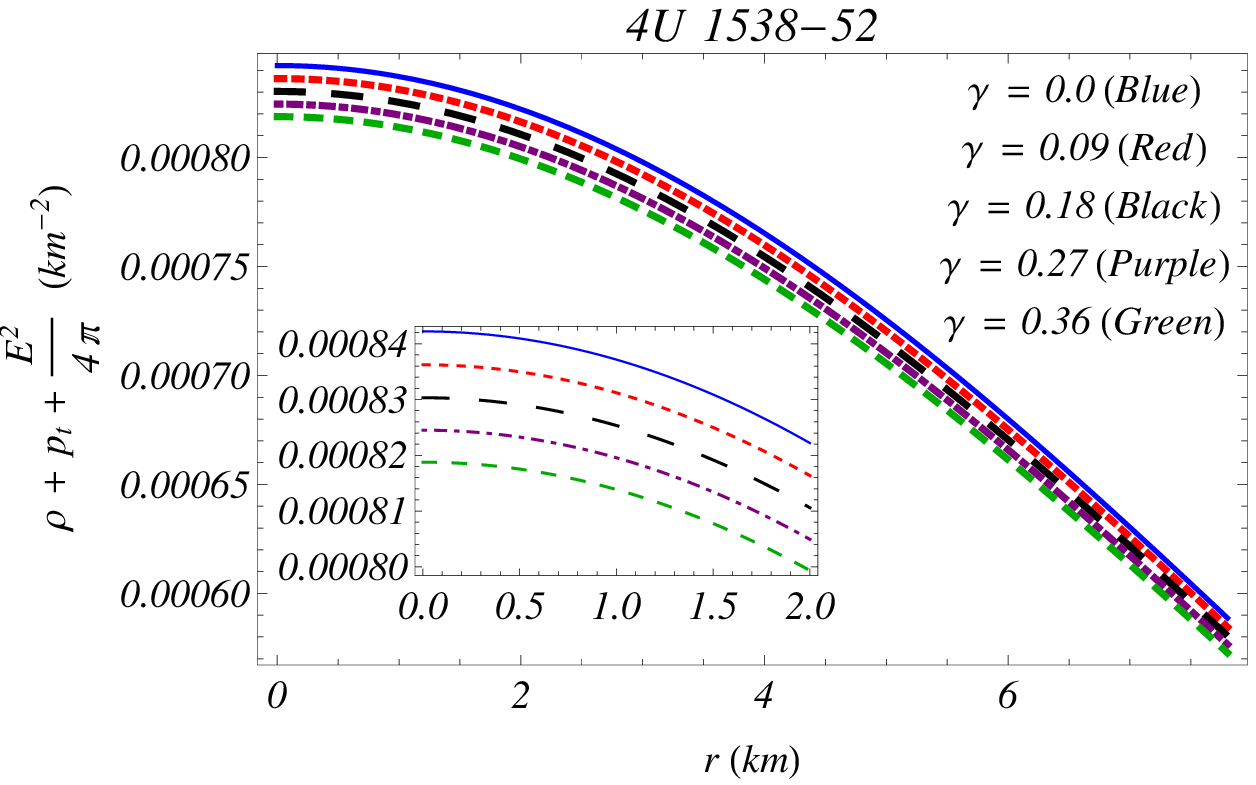}
        \includegraphics[scale=.65]{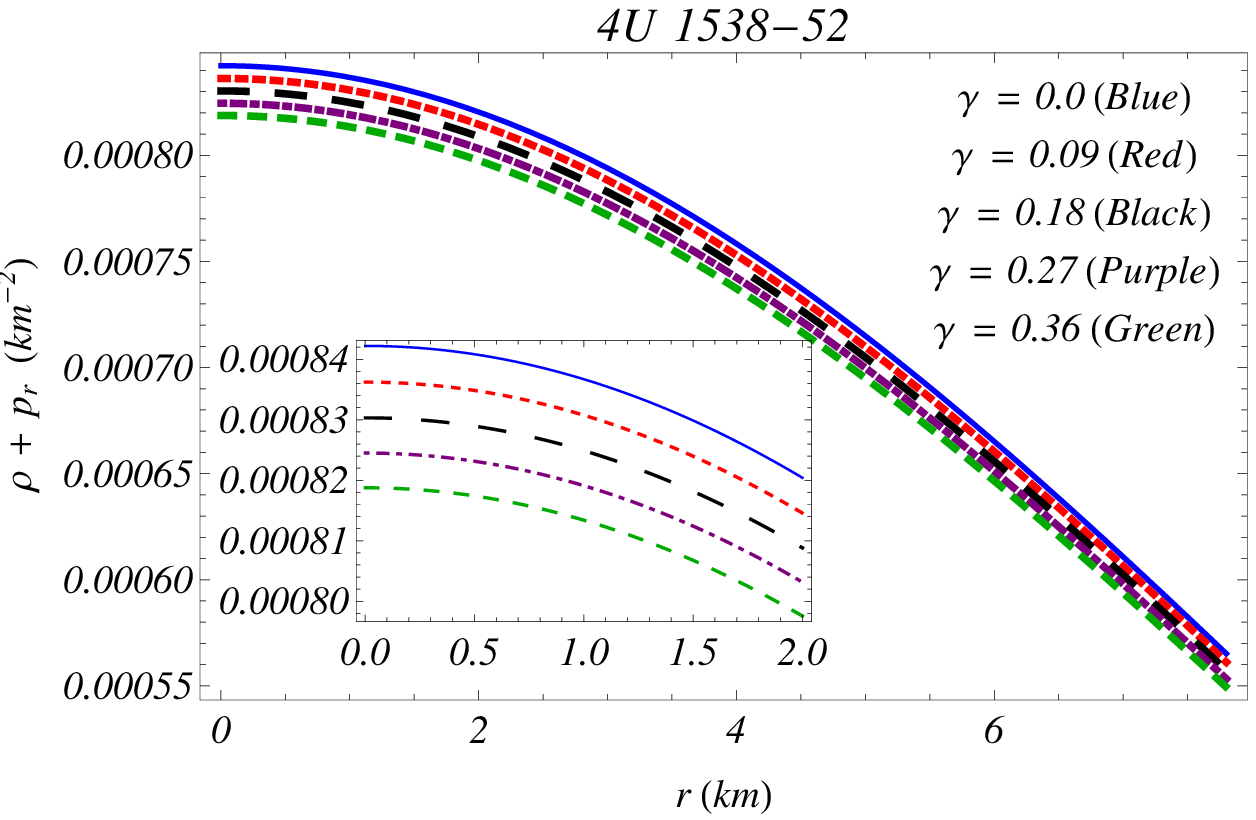}
        \includegraphics[scale=.65]{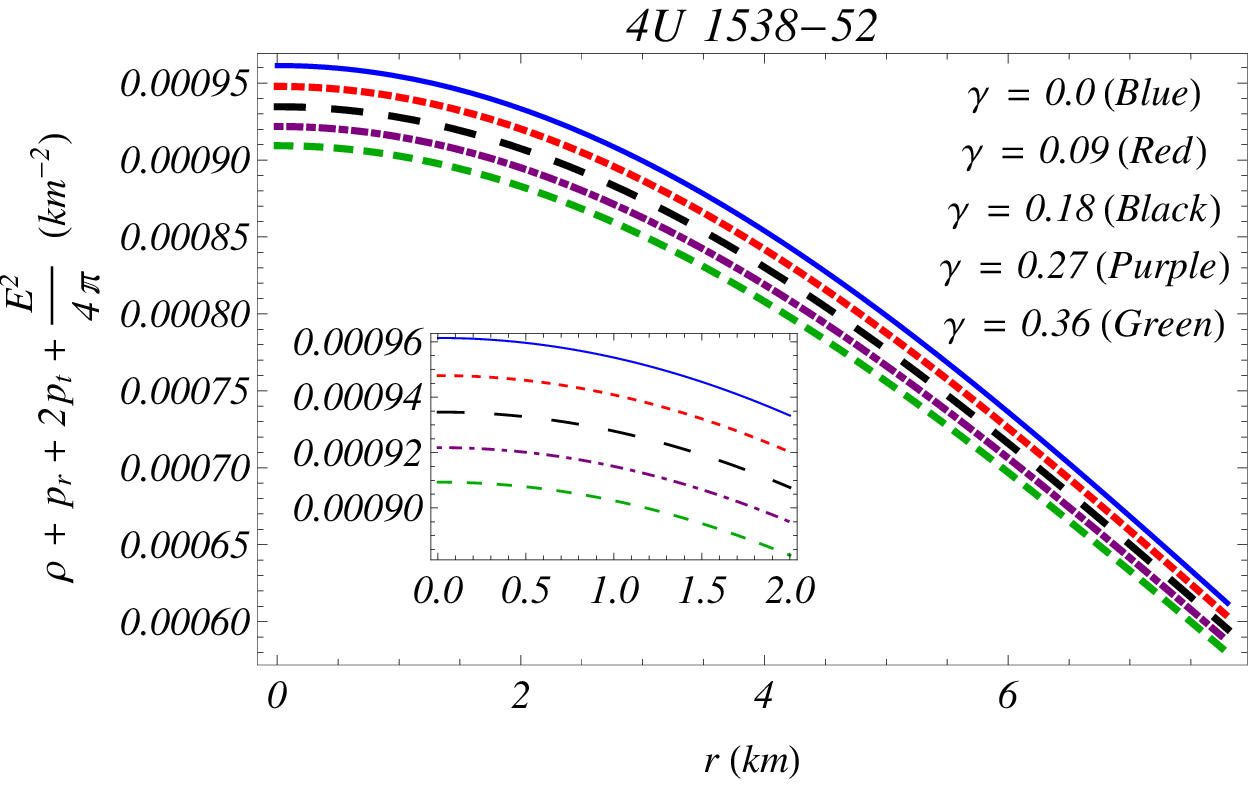}
        \includegraphics[scale=.65]{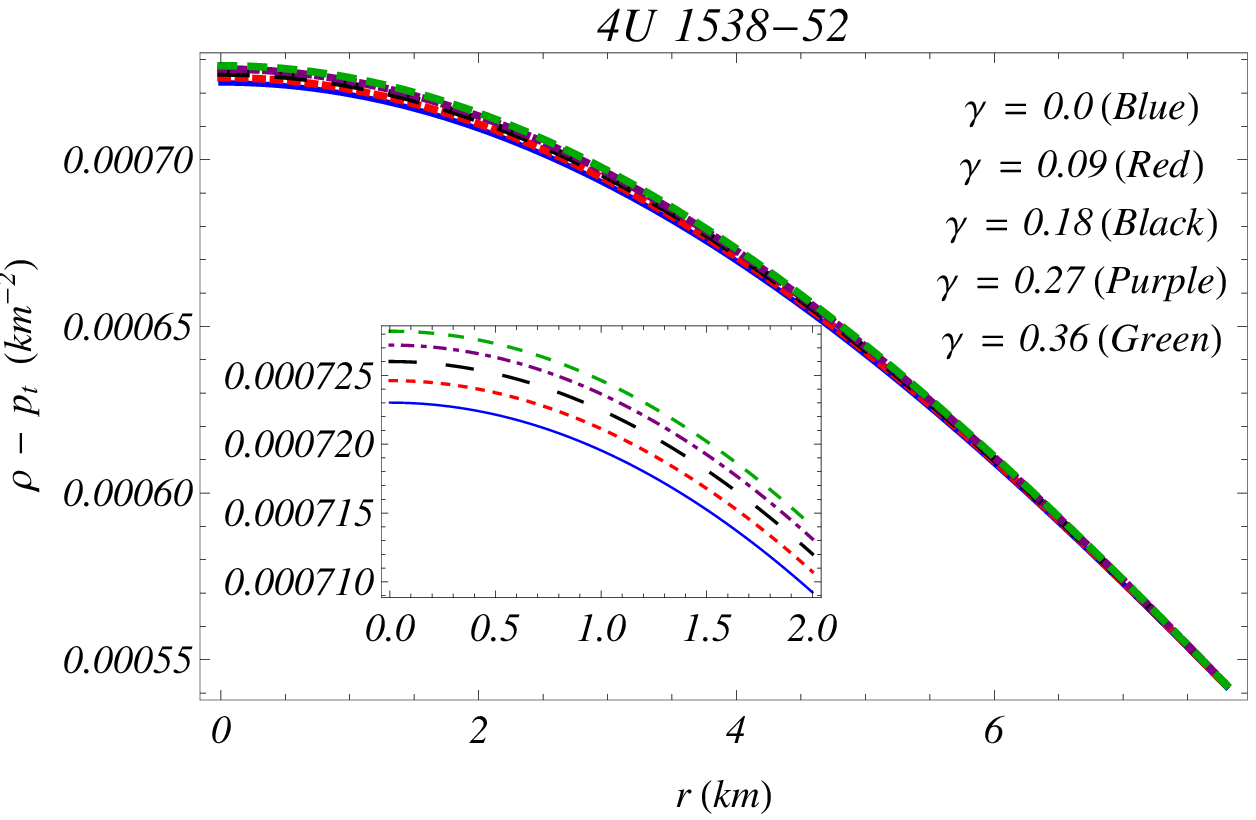}
       \caption{All the energy conditions are plotted inside the stellar interior for the strange star $4U 1538-52$ for different values of $\gamma$ mentioned in the figure.\label{ec}}
\end{figure*}

If there is the existence of exotic matter inside the compact star, the violation of the energy conditions will occur which was seen to describe the model of wormhole in the context of Einstein's general theory of relativity \cite{p101}. If these conditions are satisfied, the existence of ordinary matter is confirmed. We have checked the validity of these conditions
graphically in Fig. \ref{ec} for different values of $\gamma$ and we see that our charged anisotropic model in $f(R,T)$ gravity satisfies all the above mentioned energy conditions.

\subsection{ Pressure-density relationship}
\begin{figure*}[htbp]
    \centering
        \includegraphics[scale=.65]{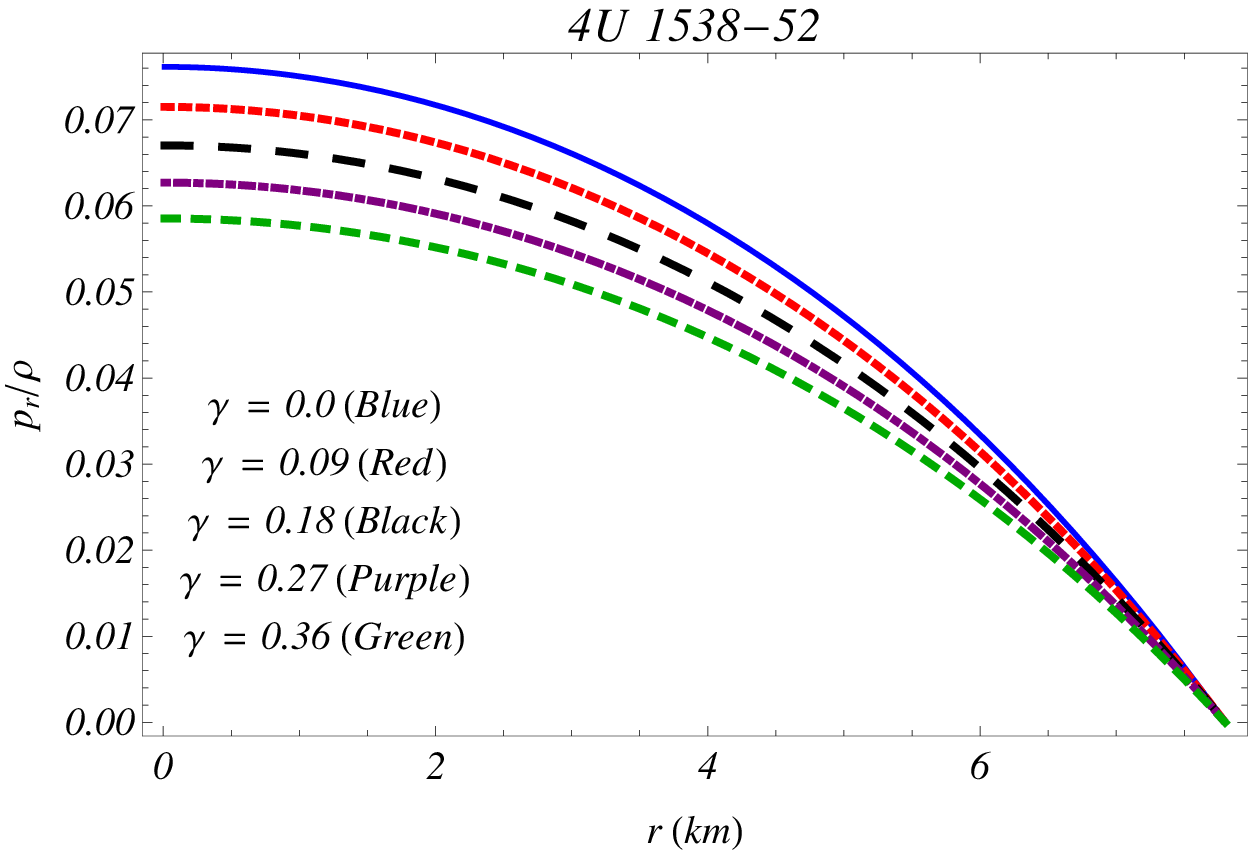}
        \includegraphics[scale=.65]{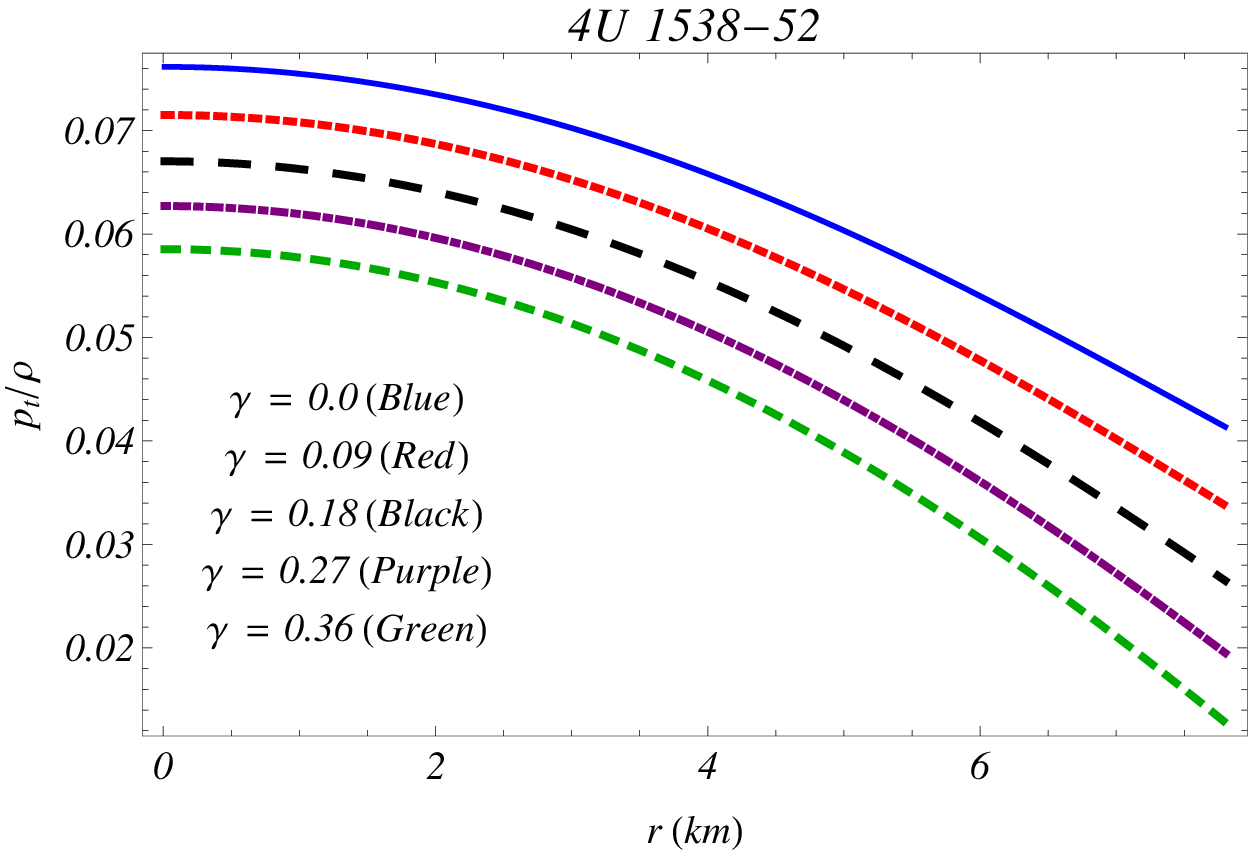}
        \includegraphics[scale=.65]{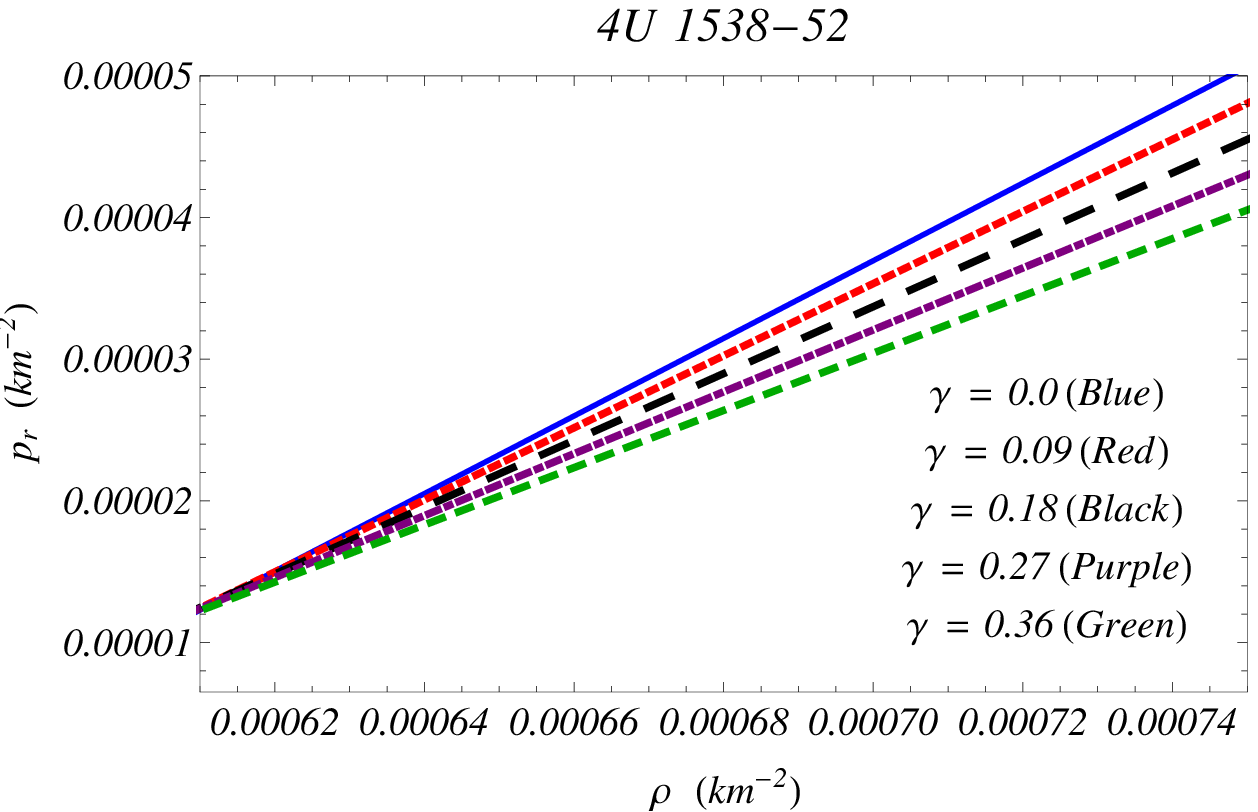}
        \includegraphics[scale=.65]{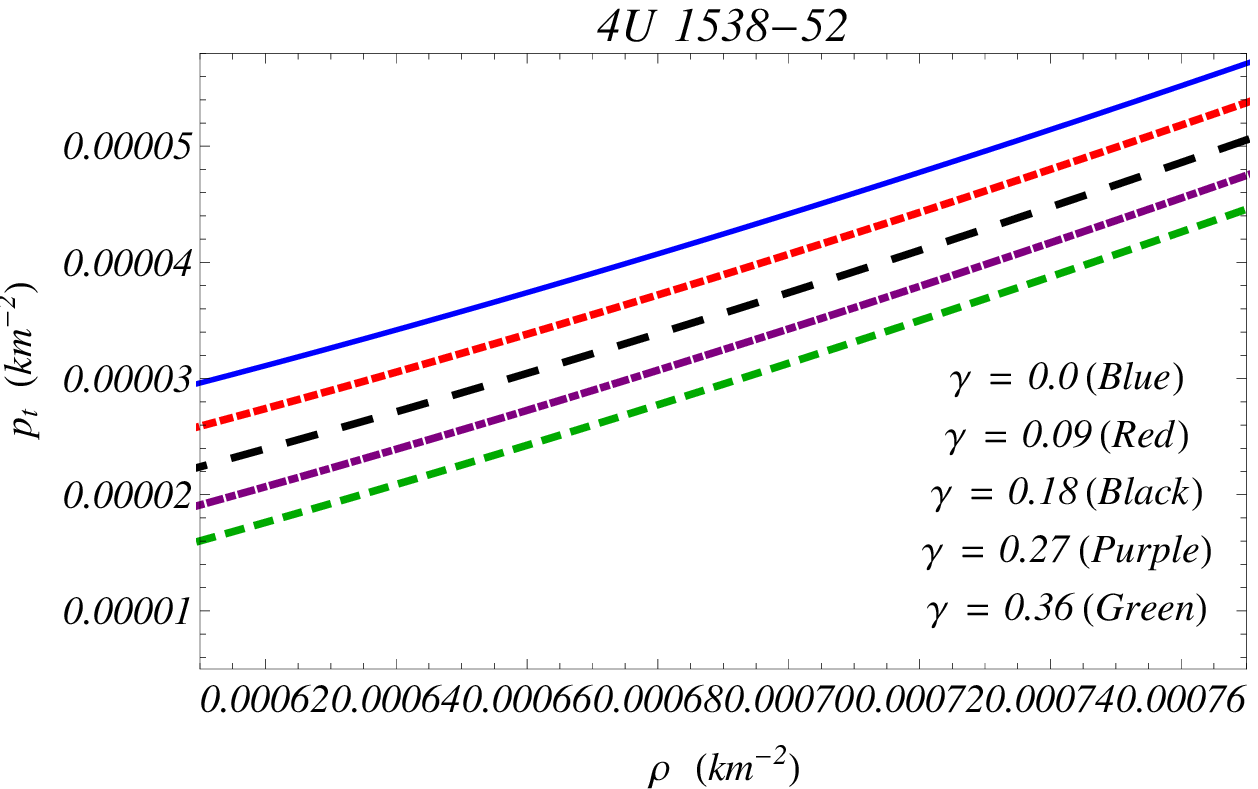}
       \caption{The variation of radial and transverse pressures are shown against matter density $\rho$ for different values of $\gamma$ mentioned in the figure.\label{eos}}
\end{figure*}

It is also important to find out a relationship between the pressure and density which is known as the equation of state. The model is developed by solving the field equations by taking a linear relationship between the radial pressure and matter density, but, the relationship between the transverse pressure and the matter density is still unknown. We have graphically shown the nature of the variation of pressure with respect to the density in Fig.~\ref{eos}.\par
The relation between matter density and pressures can be described by two dimensionless quantity which is called the equation of
state parameters and are represented by $\omega_r$ and $\omega_t$. The equations of state parameters $\omega_r$ and $\omega_t$ for our present model are obtained as,
\[p_r=\omega_r \times \rho,~~p_t=\omega_t \times \rho,\]

We have drawn the profiles of both $\omega_r,\,\omega_t$ in Fig.~\ref{eos} for different values of $\gamma$. From the figures, it is clear that these two parameters take the maximum value at the center of the star and decrease towards the boundary. Moreover, they lie in the range $0<\omega_r,\,\omega_t<1$, which corresponds to the radiation era \cite{sharif187}.

\section{Stability condition of the present model}\label{sec8}
\subsection{Equilibrium under forces }
The equilibrium of the system depends on the TOV equation \cite{t1,t2}.
Using equations (\ref{f1})-(\ref{f3}), the generalized TOV equation for our present model in $f(R,T)$ gravity can be obtained as,
\begin{eqnarray}\label{con}
-\frac{\nu'}{2}(\rho+p_r)-\frac{dp_r}{dr}+\frac{2}{r}(p_t-p_r)+\frac{\gamma}{8\pi+2\gamma}(\rho'+p_r'+2p_t')
\nonumber\\+\frac{8\pi}{8\pi+2\gamma}\frac{q}{4\pi r^4}\frac{dq}{dr}=0,
\end{eqnarray}
Using the TOV equation we want to investigate whether our present stellar system is in a stable equilibrium stage under the five following forces:~(i) the hydrostatic force $F_h$, (ii) the gravitational force $F_g$ , (iii) the anisotropic repulsive force $F_a$ , (iv) the electric force $F_e$ and (v) the force corresponding to the modified gravity $F_m$.
In eqn.(\ref{con}), for $\gamma=0$ we regain the conservation equation in Einstein gravity with the charged distribution.
\begin{figure*}[htbp]
    \centering
        \includegraphics[scale=.65]{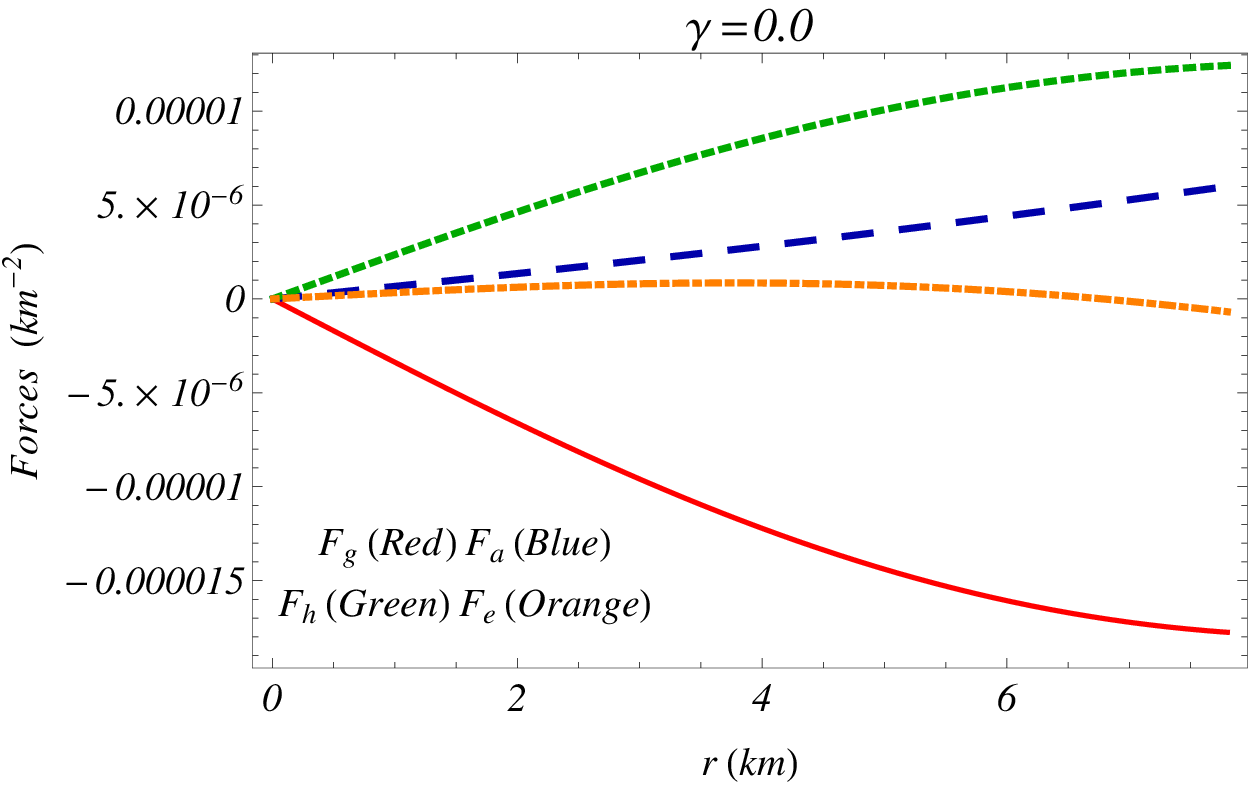}
        \includegraphics[scale=.65]{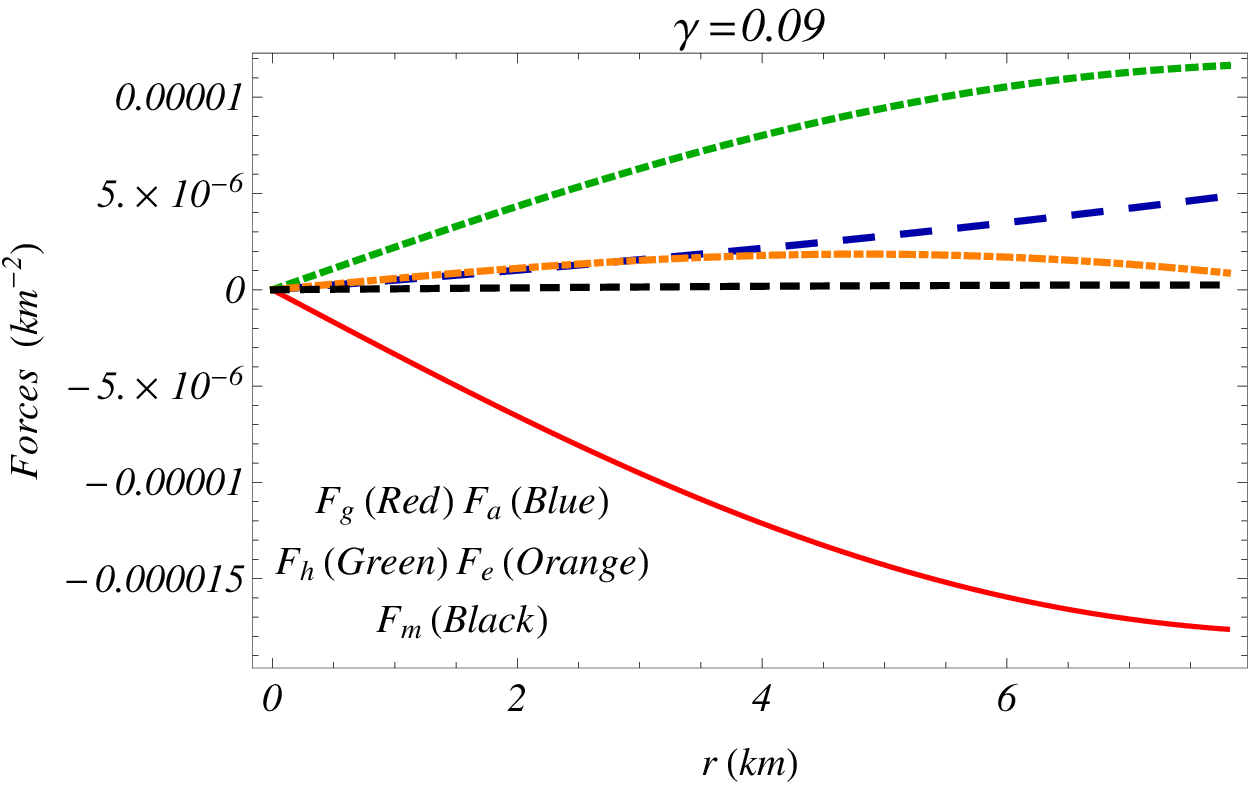}
        \includegraphics[scale=.65]{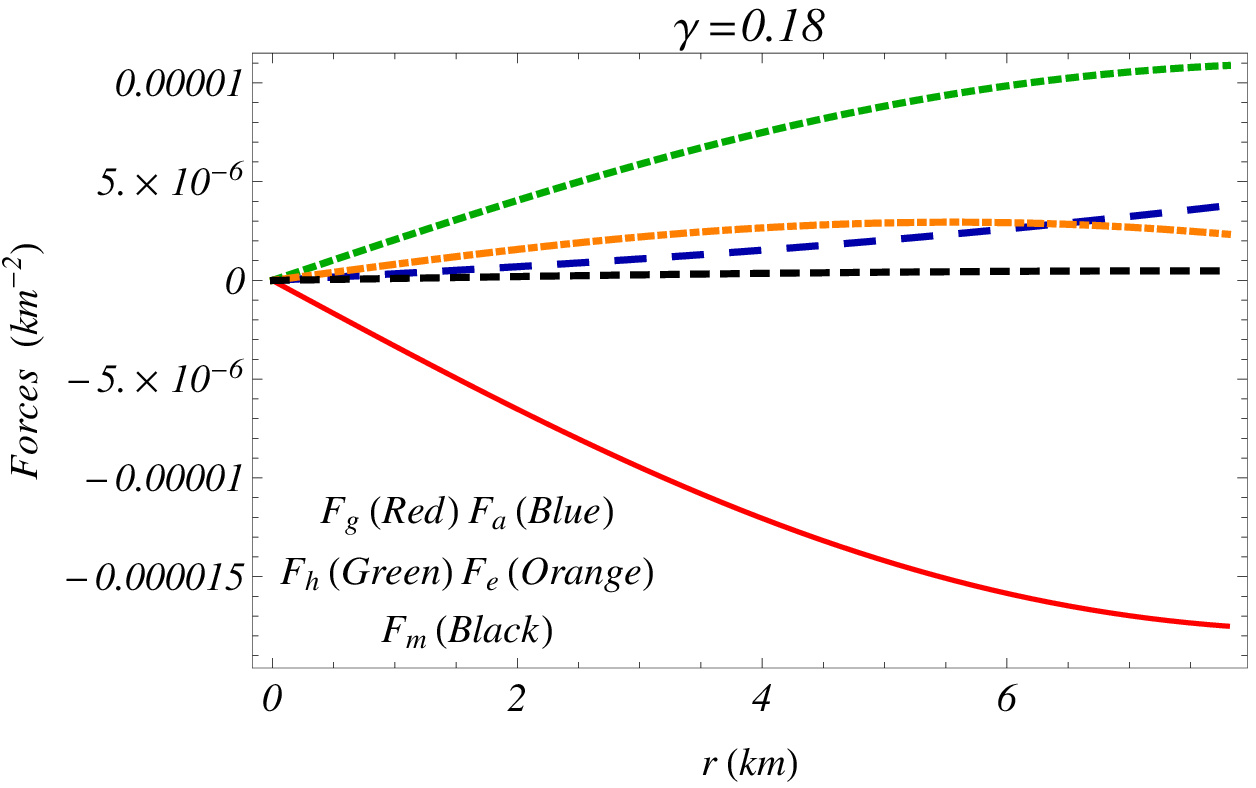}
        \includegraphics[scale=.65]{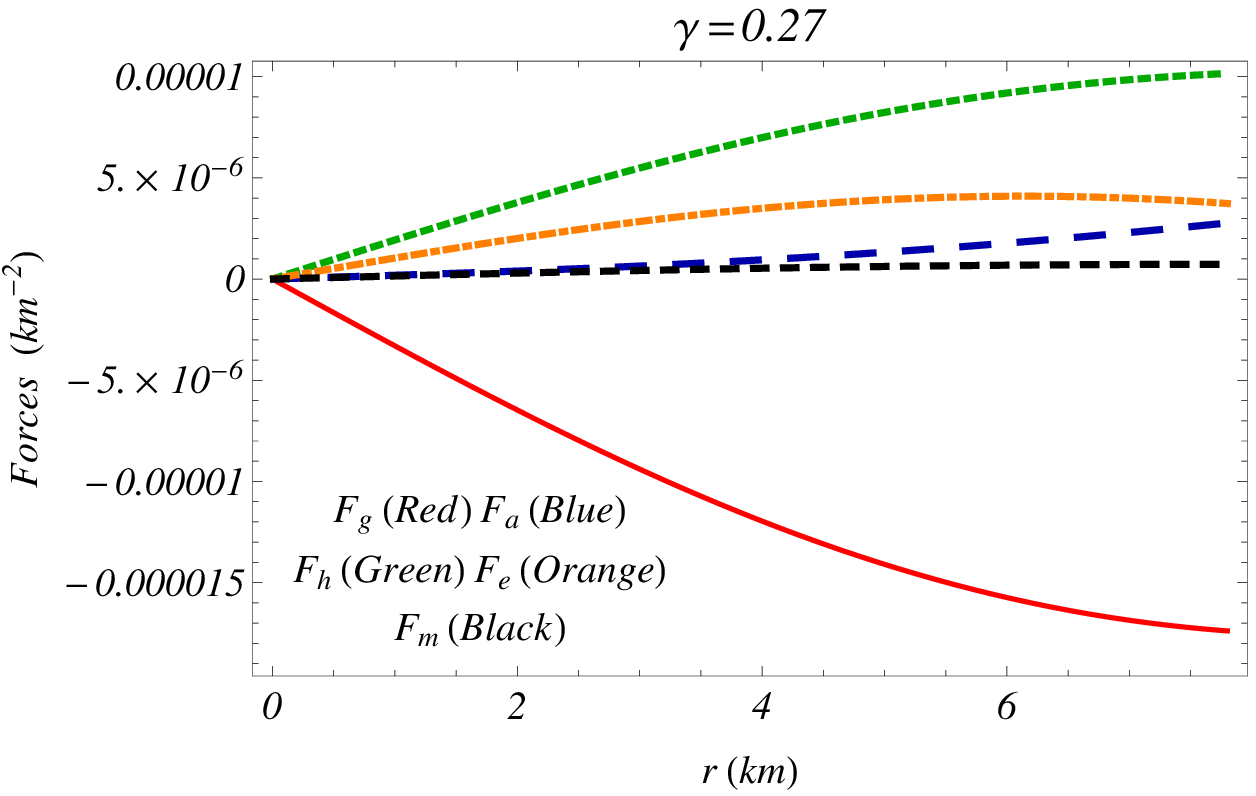}
        \includegraphics[scale=.65]{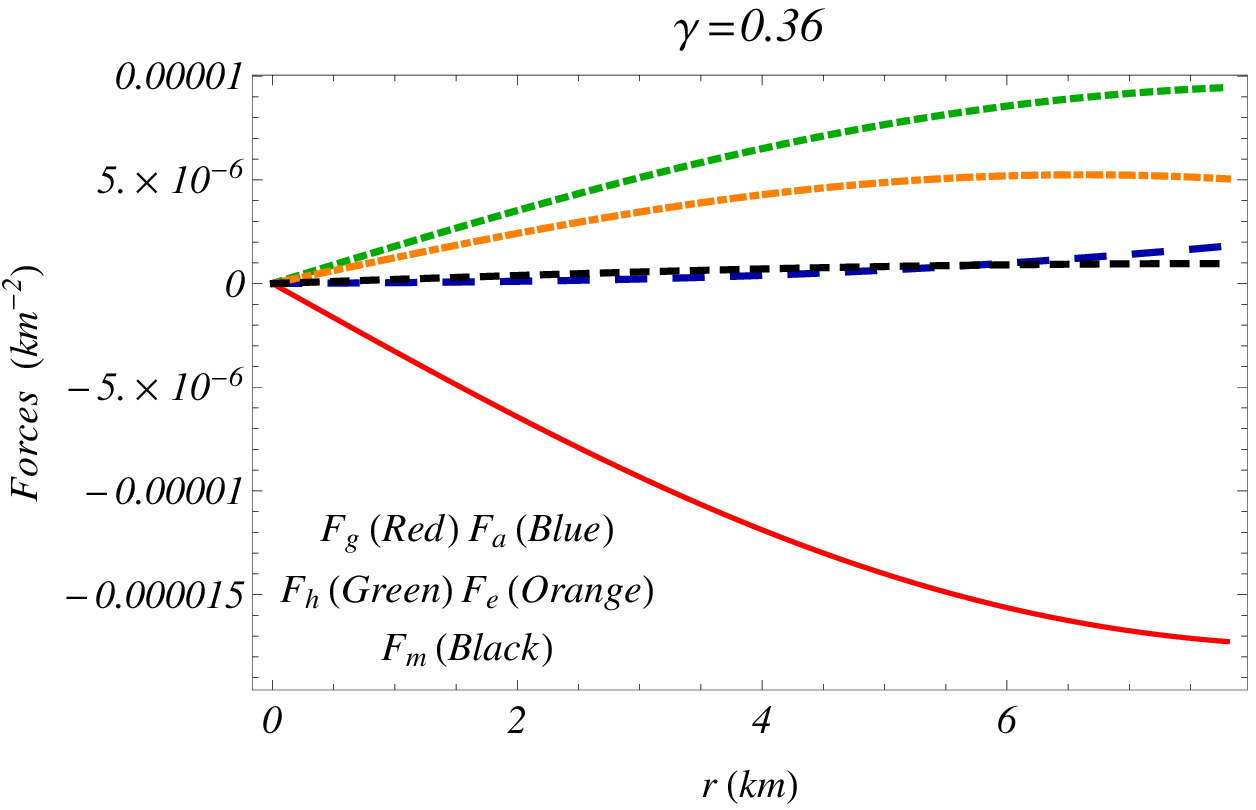}
       \caption{Different forces acting on the present model are plotted against r for the strange star candidate 4U 1538-52 by taking different values of $\gamma$.\label{tov}}
\end{figure*}

Now the expressions of different forces acting on the system are given as,
\begin{eqnarray}
F_g&=&-\frac{\nu'}{2}(\rho+p_r)\\&=&-\frac{B (A + B) e^{-A r^2} r}{\gamma + 4 \pi},\\
F_h&=&-\frac{dp_r}{dr}= \frac{2 A \alpha (A + B) e^{-A r^2} r}{(1 + \alpha) (\gamma + 4 \pi)},\\
F_a&=&\frac{2}{r}(p_t-p_r)=\frac{2}{r}\Delta \\
F_e&=&\frac{8\pi}{8\pi+2\gamma}\frac{q}{4\pi r^4}\frac{dq}{dr}\nonumber\\&=& \frac{1}{4\pi+\gamma}\left(\frac{2}{r}E^2+\frac{1}{2}\frac{d}{dr}(E^2)\right),\nonumber\\
\\
F_m&=&\frac{\gamma}{8\pi+2\gamma}(\rho'+p_r'+2p_t')\nonumber\\&&= \frac{e^{-A r^2} \gamma}{(1 + \alpha) (\gamma + 4 \pi) (3 \gamma + 4 \pi) r^3}\bigg[1 - e^{A r^2} + A r^2 \nonumber\\
&& + B r^4 \big(6 A - B + A (-A + B) r^2\big)- \alpha \Big\{e^{A r^2}\nonumber\\&&
 - \big(1-(A - B) r^2\big) \times \nonumber\\&&\big(1 + (2 A - B) r^2 + A B r^4\big)\Big\}\bigg].
\end{eqnarray}
By taking different values of $\gamma$ (mentioned in the figure), all the forces acting on the system are shown in Fig.~\ref{tov}. From the figure, one can note that the gravitational force is dominating in nature and it is attractive in nature and the effect of the force due to modified gravity is very less among all the forces. The figures verify that the combined effect of all the forces becomes zero and hence the equilibrium condition of the charge compact star in $f(R,T)$ gravity is obtained. It is worthwhile to mention that the expression for the electric field, pressure, and density gradient used in the expressions of different forces are given in eqns. (\ref{k4}),(\ref{k1})-(\ref{k3}) respectively.

\begin{figure*}[htbp]
    \centering
        \includegraphics[scale=.5]{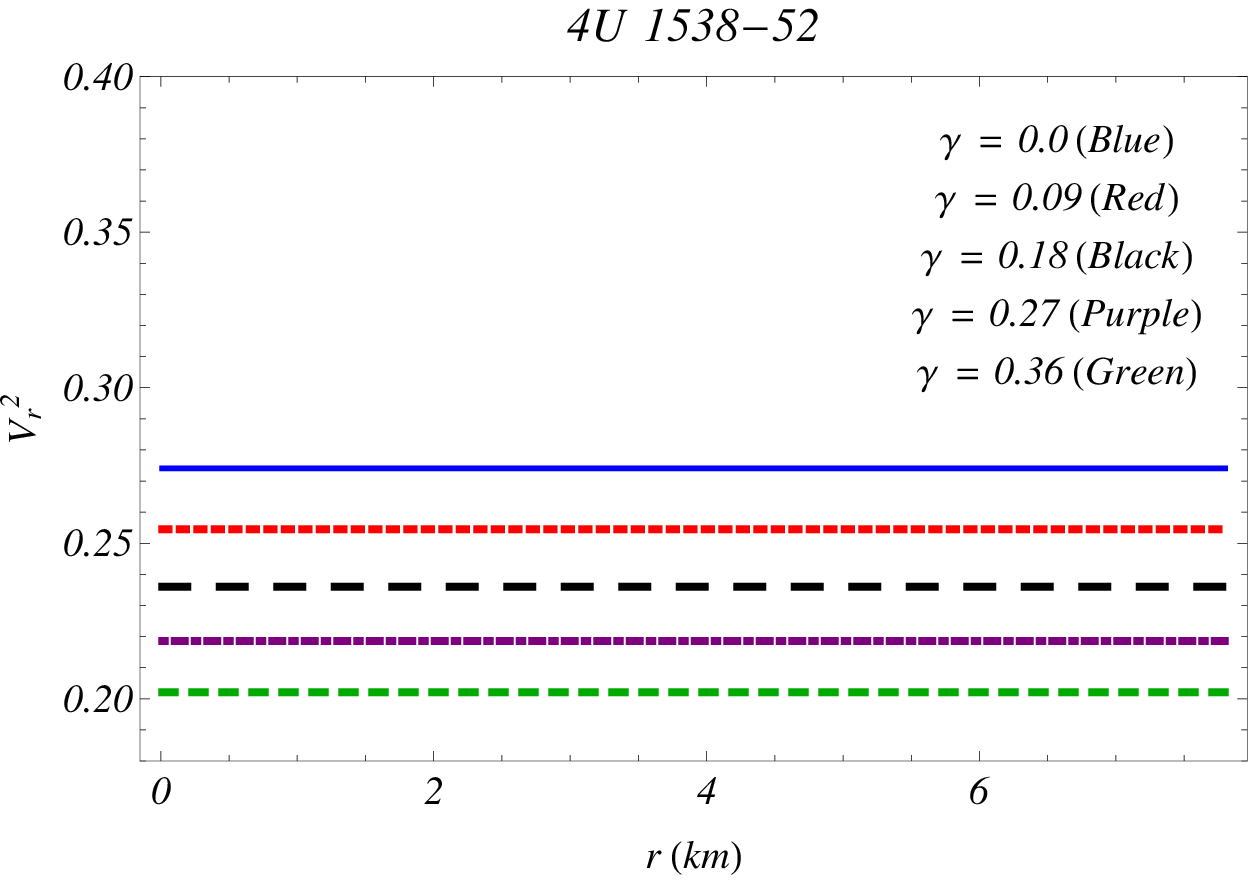}
        \includegraphics[scale=.5]{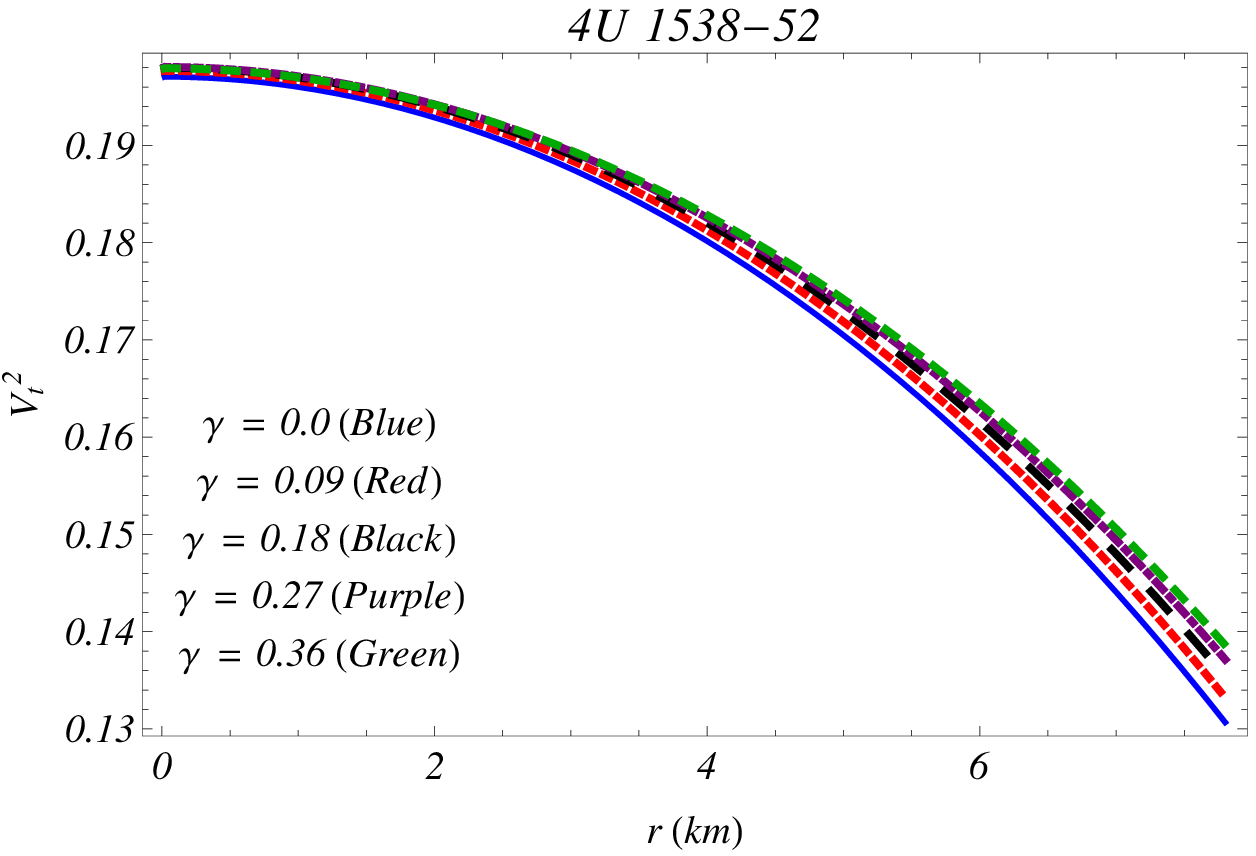}
        \includegraphics[scale=.5]{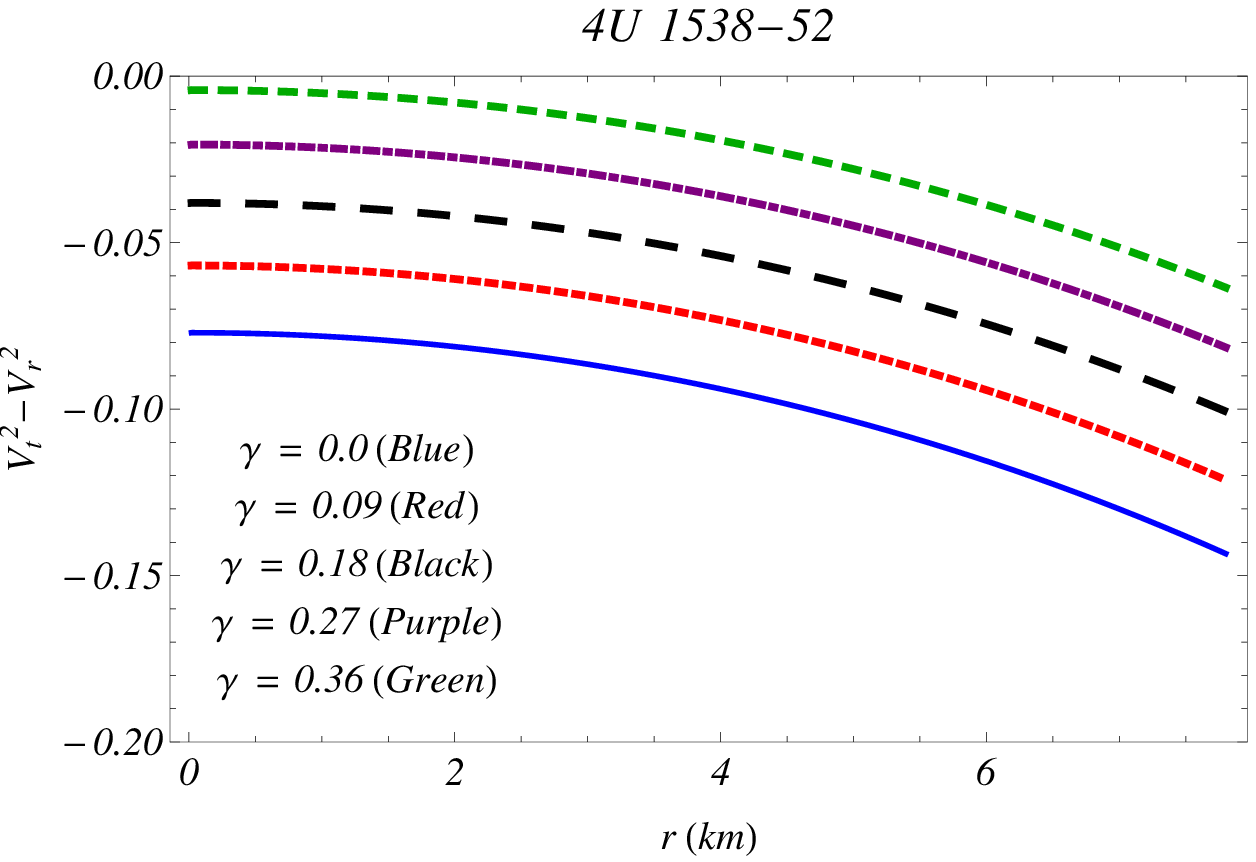}
       \caption{(Top left) Square of the radial sound velocity $V_r^2$, (top right) square of the transverse sound velocity $V_t^2$ and (bottom) the stability factor $V_t^2-V_r^2$ is plotted against r for the strange star candidate 4U 1538-52 by taking different values of $\gamma$.\label{sv}}
\end{figure*}

\subsection{Causality condition and cracking method}
The velocity of sound plays a crucial role to describe a model of a compact object which relates the pressures and density. For our present model, the square of radial and transverse speed of sounds are obtained as,
\begin{eqnarray}
V_r^2&=& \frac{dp_r}{d\rho}=\frac{p_r'}{\rho'}=\alpha,\\
 V_t^2&=& \frac{dp_t}{d\rho}=\frac{p_t'}{\rho'}\\
 &=&\frac{1}{2 A (A + B) (3 \gamma + 4 \pi) r^4}\Big[\gamma + \alpha \gamma + 4 \pi \nonumber\\&&+
  4 \alpha \pi  - (1 + \alpha) e^{A r^2} (\gamma + 4 \pi) +
  A \gamma (1+\alpha)r^2 \nonumber\\&&  + 4 A \pi r^2
  +   4 A \alpha \pi r^2 - D_1 r^4 - D_2 r^4 -D_3 r^4\nonumber\\&& -
 D_4 r^4 +D_5 r^6  +
 D_6 r^6+ D_7 r^6 + D_8 r^6\Big].
\end{eqnarray}
where, all $D_i$'s are constants and they are given by,
\[D_1=(3 A^2 - 3 A B + B^2) \gamma,~~D_2=\alpha (5 A^2 - A B + B^2) \gamma,\]\[D_3=4 (A^2 - 5 A B + B^2) \pi,~D_4= 4 \alpha (3 A^2 - 3 A B  + B^2) \pi,\]\[D_5= A B (-A + B) \gamma, ~~D_6= A \alpha B (-A + B) \gamma\]\[D_7=4 A B (-A + B) \pi ,~~D_8= 4 A \alpha B (-A + B) \pi. \]
In this study, we have taken the speed of light $c$ to be $1$, so
the stability condition for the model is $0~\leq~V_r,\,V_t~\leq~1$. The
lower bound in the previous inequality prevents dark energy fluctuations from
growing exponentially which can lead to non-physical
situations and the upper one is imposed in order to
avoid super-luminal propagation which is known as causality condition.

\begin{figure}[htbp]
    \centering
        \includegraphics[scale=.6]{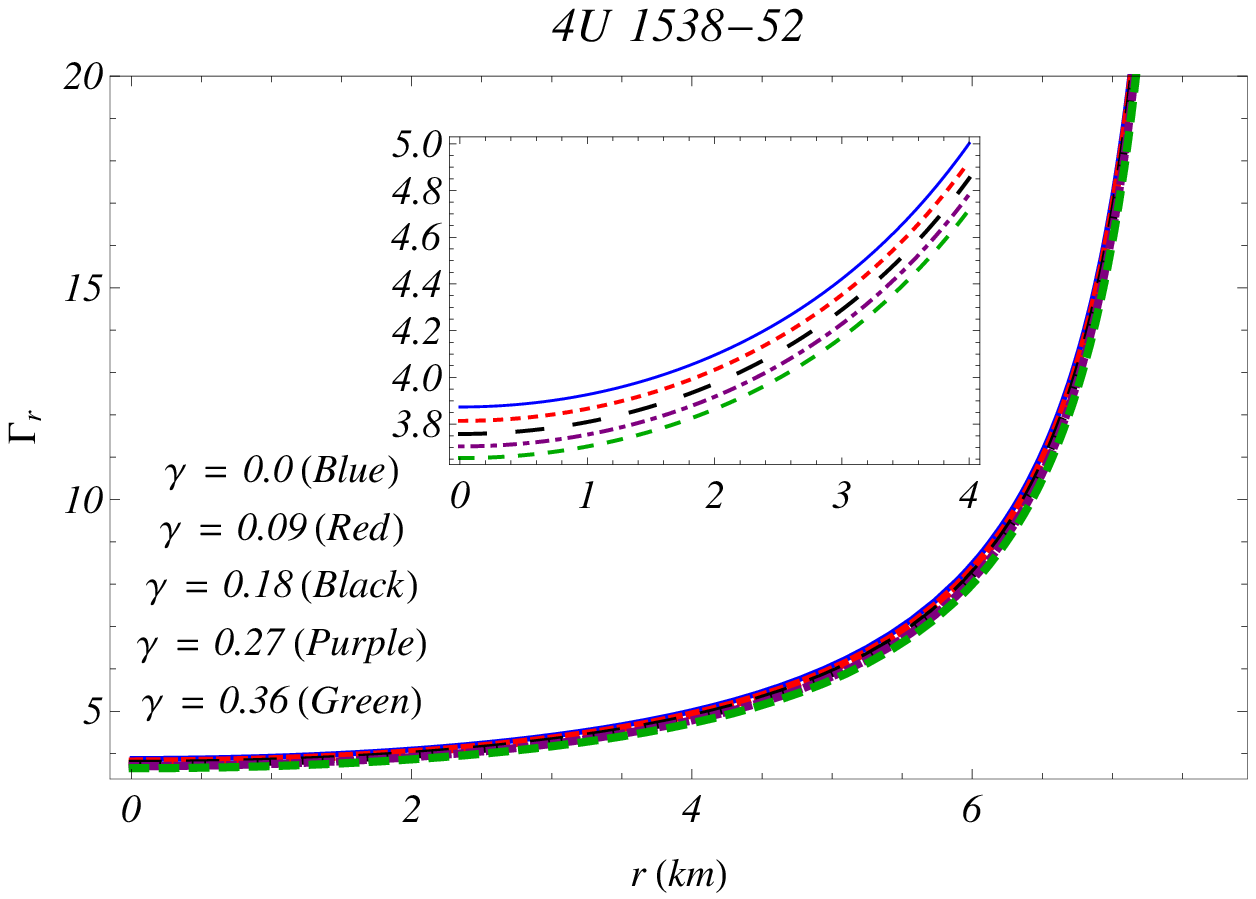}
        \includegraphics[scale=.6]{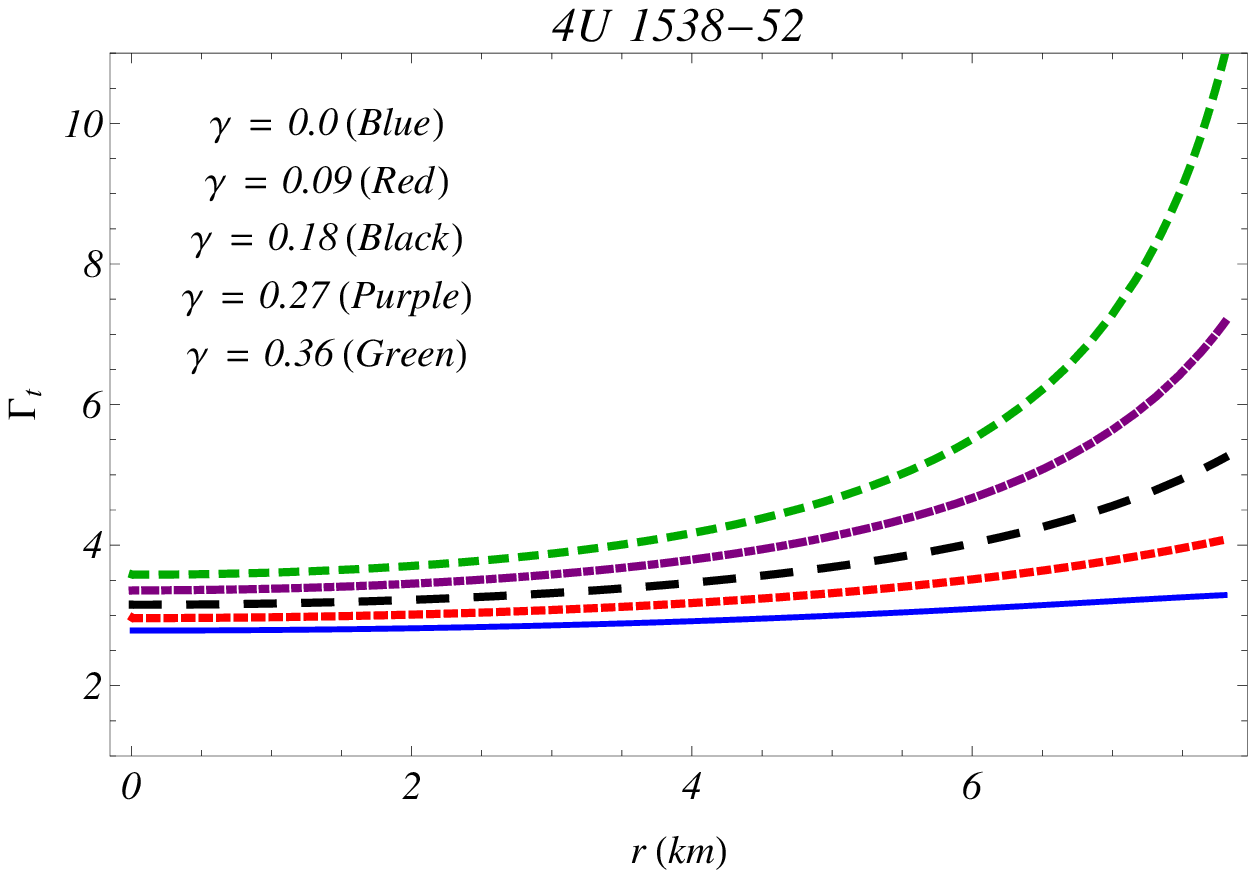}
       \caption{The relativistic adiabatic index $\Gamma_r$ and $\Gamma_t$ have been plotted against r inside the stellar interior }\label{fgh1}
\end{figure}

According to the graph shown in Fig.~\ref{sv}, it can be seen that the
model satisfies $V_r^2,~V_t^2~\in~(0,\,1)$ everywhere inside the boundary. Therefore, we can
say that our model satisfies causality condition.\\
In a series of lectures \cite{a10,a11,a12}, Herrera and collaborators discussed in detail the concept of cracking for self-gravitating isotropic and anisotropic matter configurations. To identify the potentially unstable anisotropic matter
configurations, in 1992, \cite{a10} introduced the concept of cracking (or overturning) which is indeed a very useful technique. Later, Herrera along with his collaborators showed that even small deviations from local isotropy may lead to drastic changes in the
evolution of the system as compared with the purely locally isotropic case. Now, we have,
\[V_t^2-V_r^2~\sim~\frac{dp_t}{d\rho}-\frac{dp_r}{d\rho}~\sim~\frac{d(p_t-p_r)}{d\rho}~\sim~ \frac{d \Delta}{d \rho}. \]
Now since $V_r^2,\,V_t^2~\in~(0,\,1)$ it implies, $-1~<V_t^2-V_r^2~<1$.
Which further gives, \[|V_t^2-V_r^2|~\leq~1~\rightarrow~|\frac{d \Delta}{d \rho}|~\leq~1~\rightarrow~|d \Delta|\leq |d \rho|.\] These perturbations lead to potentially unstable models when $\frac{d \Delta}{d\rho}>0$, so the potential stability region is therefore obtained from the inequality $V_t^2-V_r^2~<0$.\\
The profiles of $V_t^2-V_r^2$ for different values of $\gamma$ have been shown in Fig.~\ref{sv}. The profiles ensure that our model is potentially stable since $V_t^2-V_r^2~<0$ holds for all $r~\in~[0,\,R].$

\begin{figure*}[htbp]
    \centering
        \includegraphics[scale=.6]{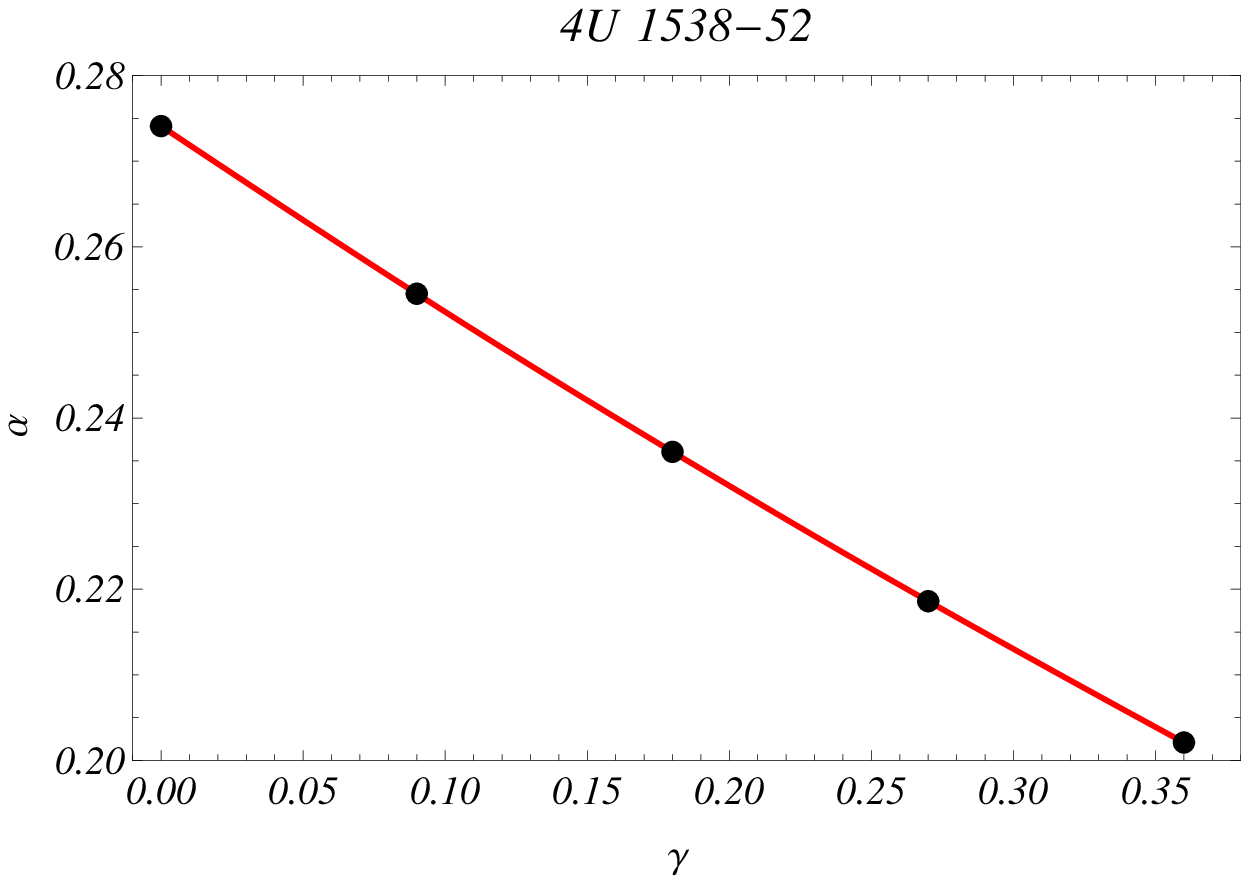}
        \includegraphics[scale=.6]{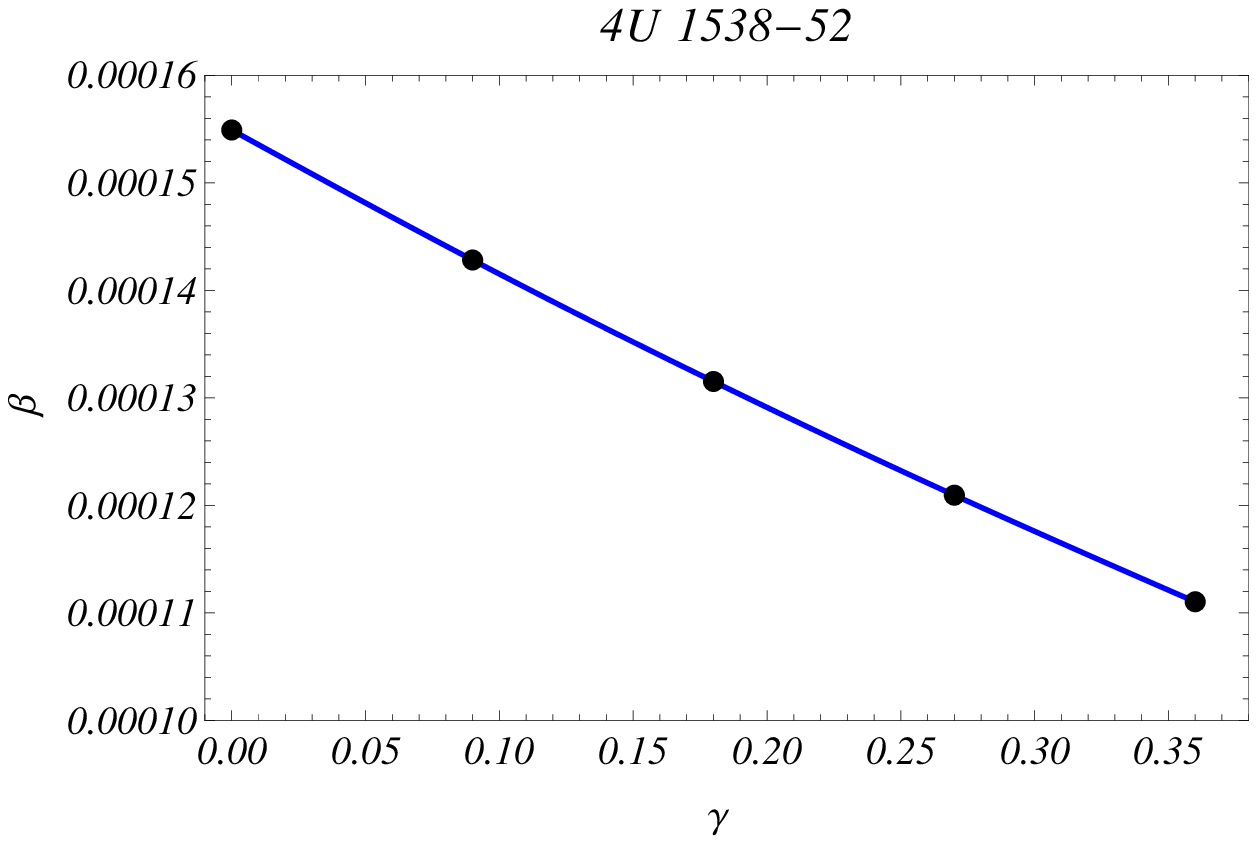}
        \includegraphics[scale=.6]{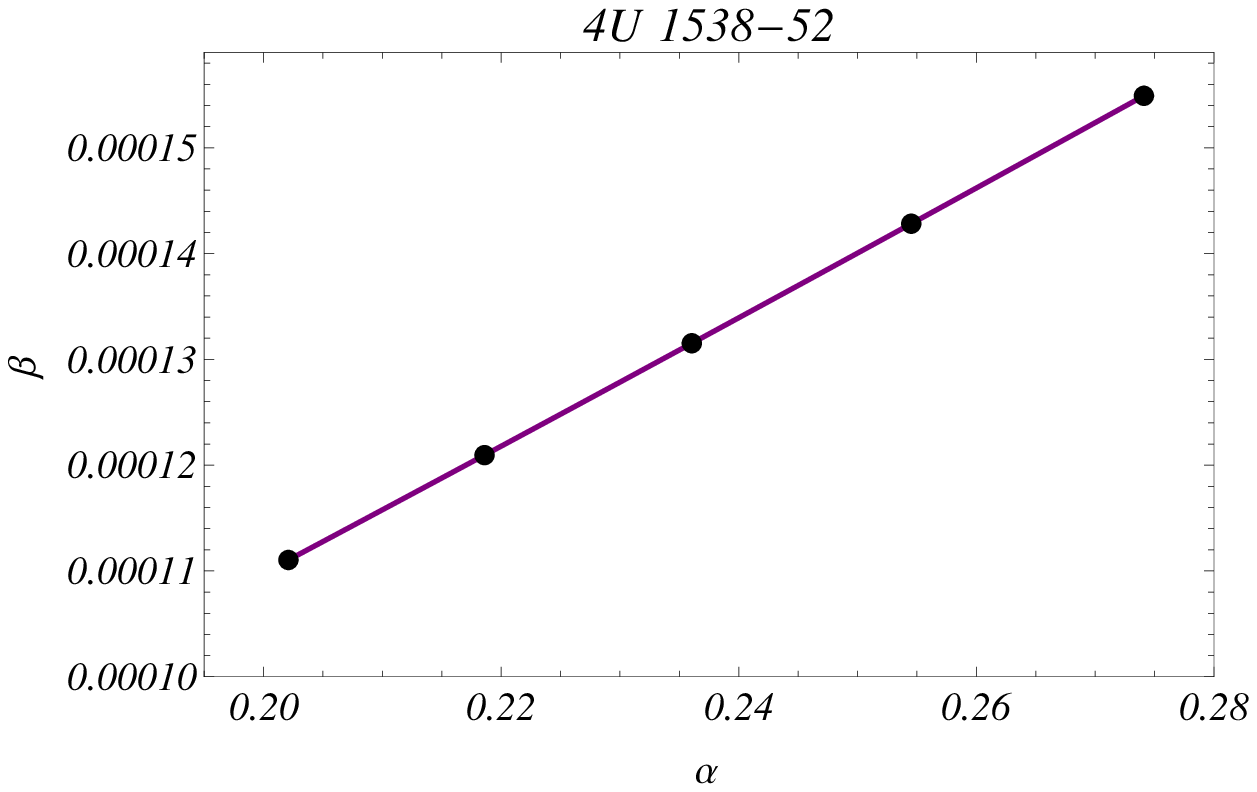}
       \caption{The variations of (top left) $\alpha$ with respect to $\gamma$, (top right) $\beta$ with respect to $\gamma$ and (bottom) $\beta$ with respect to $\alpha$ have been depicted.\label{par}}
\end{figure*}

\subsection{Relativistic Adiabatic index }
In this subsection, we shall explore an important and necessary ratio of two specific heats given by $\Gamma$ to examine the region of stability of spherical stars. The adiabatic index $\Gamma$ for an isotropic fluid sphere was proposed
by \cite{chan10} as, $\Gamma=\frac{\rho+p}{p}\frac{dp}{d\rho}$, but, \cite{255} was the pioneer in this era who explored the role of the adiabatic index to examine the region of stability for spherical stars.  In presence of pressure anisotropy $\Delta=p_t-p_r$ the expression for $\Gamma$ changes as,
\begin{eqnarray}
\Gamma_r=\left(\frac{\rho}{p_r}+1\right)\frac{dp_r}{d\rho},~~\Gamma_t=\left(\frac{\rho}{p_t}+1\right)\frac{dp_t}{d\rho}.
\end{eqnarray}
\cite{bondi25} pointed out that for the instability of the model $\Gamma<\frac{4}{3}$. \cite{hil} examined
that when $\Gamma$ is higher than $4/3$ and in the presence of a growing, positive
anisotropy factor, the relativistic stability condition is satisfied by the compact object. We have drawn the profiles of $\Gamma_r,\,\Gamma_t$ for different values of $\gamma$ in Fig.~\ref{fgh1}, since it is not possible to check this criterion for the complexity of the expression of $\Gamma_r$ and $\Gamma_t$. The figure indicates that both $\Gamma_i$ ($i=t,\,r$) takes the values more than $4/3$ everywhere inside the fluid sphere and it ensures that stability condition is well satisfied.

\section{Discussion}

We have investigated a new family of solutions of a charged compact star in the context of $f(R,T)$ modified theory of gravity based on linear functional forms of $f(R, T)$ as $f(R,T)=R +2 \gamma T$ where $R$ and $T$ are respectively the Ricci scalar and the trace of energy-momentum tensor respectively. To solve the field equations we have three equations with six unknowns. To solve the system, we have to choose any three of them and by our knowledge of school level Algebra, we can perform it in ${6 \choose 3}=20$ ways. Our system is governed by the linear equation of state given by $p_r=\alpha \rho-\beta$, where $\alpha$ and $\beta$ are positive constants and they depend on the coupling constant $\gamma$. The linear equation of state is a generalized version of MIT bag model EOS $p_r=(\rho-4B_g)/3$, where $B_g$ is the bag constant. For the present study, we have
taken spherically symmetric geometry with the matter contents involving anisotropic pressure profile
with a net electric charge. Here we have assumed $\nu = Br^2+C$ and $\lambda=Ar^2$ for $g_{tt}$ and $g_{rr}$ metric
components as proposed by \cite{kb}. In our present analysis, we have chosen $\gamma$ as a small positive constant since negative values of $\gamma$ produce a negative value of the electric field $E^2$. For our graphical analysis, we have considered the star $4U 1538-52$ whose mass and radius is measured by $(0.87 \pm 0.07)~M_{\odot}$ and $7.866 \pm 0.21$ km. respectively \cite{mus}. $4U 1538-52$ was discovered by the Uhuru satellite \cite{g1} is an X-ray pulsar with a B-type supergiant companion, QV Nor. It has an orbital period of $\sim~3.728$ days \cite{dav1,dav2} with eclipses
lasting $\sim~0.6$ days \cite{be}. $4U 1538-52$ was observed with the XMM-Newton satellite in
2003 from August 14 15:34:01 to August 15 14:02:30 UT using three European Photon Imaging
Cameras (EPIC)/PN and EPIC/MOS. In $4U 1538-52$ the emission line at $\sim~6.4 keV$ can usually
be described within the uncertainties either by a single narrow Gaussian line or by multiples of narrow Gaussian lines \cite{white}. The profile of both the metric potential $e^{\lambda}$ and $e^{\nu}$ are plotted in Fig.~\ref{metric}. A smooth matching of the metric potentials to the exterior spacetime is also shown in that figure.
For the coupling constant $\gamma=0,\,0.09,\,0.18,\,0.27$ and $0.36$ the graphical illustration of different physical properties have been presented in Figures \ref{mass}-\ref{par} for the compact star $4U 1538-52$. The $\gamma=0$ corresponds to the General Relativity case. The behavior of $\alpha$ and $\beta$ with respect to $\gamma$ have been shown in Fig.~\ref{par}. From the profile, it is clear that both $\alpha$ and $\beta$ decreases if $\gamma$ increases but $\beta$ increases with the increasing value of $\alpha$. From our discussion, we see that the cental value of the matter density, as well as the surface density, decreases with the increasing value of $\gamma$. Both the radial and transverse pressure show the same behavior as the matter density. In the case of pressure anisotropy, the surface value of $\Delta$ decreases with the increasing value of $\gamma$ and $\Delta>0$ everywhere inside the fluid distribution. The positive value of the anisotropic factor creates a repulsive force which helps the stellar configuration from gravitational collapsing. On the other hand, in the case of electric field $E^2$, we see that the surface value of $E^2$ increases with the increasing value of $\gamma$. Both the pressure and density have a maximum value at the center and they decrease towards the boundary and at the boundary of the star the radial pressure vanishes at the boundary of the star, but both transverse pressure and anisotropic factor do not vanish at the boundary. We have also obtained the numerical values of the central density, surface density, and central pressure in the order of $10^{15}$~gm.cm$^{-3}$, $10^{14}$~gm.cm$^{-3}$ and $10^{34}$~dyne.cm$^{-2}$ respectively. The numerical values of the effective mass and surface redshift increases as $\gamma$ increases. Since for large values of coupling constant $\gamma$, the effective mass function increases, whereas the radius of the star is fixed, so one can conclude that the star becomes more massive, i.e., it has a trend of a higher `packing' of mass within the radius. We have obtained the numerical value of the effective surface redshift from our model for different values of $\gamma$ and we see that $z_s <0.24$ for $\gamma=0.0,\,0.09,\,0.18,\,0.27$. We found in the literature that in the absence of a cosmological
constant the surface redshift $z_s$ lies in the range
$z_s \leq 2$ \cite{buch,z2}. On the other hand, \cite{z3} showed that for an anisotropic star in the presence
of a cosmological constant the surface redshift obeys the
inequality $z_s \leq 5$. So the value of the surface redshift lies in the expected range. The stability conditions of the present model are tested via different physical conditions. With the help of graphical representation, all the energy conditions are also verified for our model.
In conclusion, we can say that, in
the current paper, we have effectively exhibited a stable and
physically adequate anisotropic astrophysical model in the background of $f(R,T)$ modified theory of gravitation which
is reasonable to study ultra-dense strange star.\\

{\bf ACKNOWLEDGMENTS:} PB is thankful to IUCAA, government of India for providing visiting associateship.

\end{document}